\documentclass[12pt]{article}
\usepackage{amssymb}

\usepackage{amsmath}
\usepackage{graphicx}

\numberwithin{equation}{section}
\providecommand{\U}[1]{\protect\rule{.1in}{.1in}}
\providecommand{\U}[1]{\protect\rule{.1in}{.1in}}
\newtheorem{theorem}{Theorem}
\newtheorem{acknowledgement}[theorem]{Acknowledgement}

\input{tcilatex}

\begin{document}

\title{About dual one-dimensional oscillator and Coulomb-like theories.}
\author{I.V. Tyutin\thanks{%
Lebedev Physical Institute. Moscow. Russia; e-mail:tyutin@lpi.ru}, G.V.
Grigoryan\thanks{%
Yerevan Physics Institute. Yerevan. Armenia; e-mail:gagri@yerphi.am}, R.P.
Grigoryan\thanks{%
Yerevan Physics Institute. Yerevan. Armenia; e-mail:rogri@yerphi.am}}
\date{ }
\maketitle

\begin{abstract}
We present a mathematically rigorous quantum-mechanical treatment of a

one-dimensional nonrelativistic quantum dual theories (with oscillator and

Coulomb like potentials) and compare their spectra and the sets of
eigenfunctions.

We construct all

self-adjoint Schrodinger operators for these theories and represent rigorous

solutions of the corresponding spectral problems. Solving the first part of

the problem, we use a method of specifying s.a. extensions by (asymptotic)

s.a. boundary conditions. Solving spectral problems, we follow the Krein's

method of guiding functionals. We show, that there is one to one

correspondence between the spectral points of dual theories in the planes

energy-coupling constants not only for discrete, but also for continuous
spectra.
\end{abstract}

\section{Introduction}

It is well known \cite{Ter-Ant}, that if one introduces in a radial part of
the $D$ dimensional oscillator ($D > 2$)

\begin{equation}  \label{eq1}
{\frac{{d^{2}R}}{{du^{2}}}} + {\frac{{D - 1}}{{u}}}{\frac{{dR}}{{du}}} - {%
\frac{{L\left( {L + D - 2} \right)}}{{u^{2}}}}R + {\frac{{2\mu} }{{\hbar ^{2}%
}}}\left( {E - {\frac{{\mu \omega ^{2}u^{2}}}{{2}}}} \right)R = 0
\end{equation}
(here $R$ is the radial part of the wave function for the $D$ dimensional
oscillator ($D>2$) and $L=0,\mathrm{1},\mathrm{2},...$ are the eigenvalues
of the global angular momentum ) $r=u^{2}$ then equation (\ref{eq1})
transforms into

\begin{equation}
{\frac{{d^{2}R}}{{dr^{2}}}}+{\frac{{d-1}}{{r}}}{\frac{{dR}}{{dr}}}-{\frac{{%
l\left( {l+d-2}\right) }}{{r^{2}}}}R+{\frac{{2\mu }}{{\hbar ^{2}}}}\left( {%
\mathcal{E}+{\frac{{\alpha }}{{r}}}}\right) R=0  \label{eq2}
\end{equation}%
where $d=D/2+1$, $l=L/2$, $\mathcal{E}=-{\frac{{\mu \omega ^{2}}}{{8}}}$, $%
\alpha =E/4$, which formally is identical to the radial equation for $d$%
-dimensional hydrogen atom.

Equations (\ref{eq1}) and (\ref{eq2}) are dual to each other and the duality
transformation is $r=u^{2}$ For discreet spectrum of these equations it was
proved, that to each state of equation (\ref{eq1}) corresponds a state in (%
\ref{eq2}), and visa versa \cite{Ners-Ter-Ant,Hak-Ter-Ant}. However the
correspondence of the states in general (for discrete, as well as continuous
spectra and for all values of the parameters of the theory) the problems was
not considered. In this article we will consider the problem for the
one-dimensional case, in which the Schr\"{o}dinger equation for the
oscillator is

\begin{equation}
{\frac{{d^{2}\Psi }}{{du^{2}}}}+\left( \frac{{2\mu }}{{\hbar ^{2}}}{E}_{O}{%
-\lambda u^{2}}\right) \Psi =0,\;\lambda =\frac{{\mu }^{2}{\omega ^{2}}}{{%
\hbar ^{2}}}  \label{eq3}
\end{equation}%
which under duality transformation $x=u|u|$ and setting 
\begin{equation}
\Psi =x^{-1/4}\Phi  \label{eq5}
\end{equation}%
transforms into 
\begin{equation}
{\frac{{d^{2}\Phi }}{{dx^{2}}}}+\left( \frac{{2\mu }}{{\hbar ^{2}}}E_{C}-{{%
\frac{{g}}{{|x|}}}+{\frac{{3}}{{16x^{2}}}}}\right) \Phi =0,  \label{eq4}
\end{equation}%
where $E_{C}$ and $g$ are some functions of parameters ${E}_{O}$ and ${%
\lambda }$.

Eq.(\ref{eq4}) includes a Coulomb-like potential and describes the so called 
$1D$ anyon. Unlike the Eq.(\ref{eq3}), which is defined for all values of
the variable, eq.(\ref{eq4}) is defined on the axis with punctured zero
point. Taking into account, that duality transformation $x=u|u|$ is also
singular at the origin, we will consider the oscillator problem also with
punctured zero point.

We will solve the quantum problem of these two equation and will show a
complete correspondence of the states for all values of the parameters ${E}%
_{O}$, ${\lambda }$, $E_{C}$, and $g$. In section 2 we will consider the
quantum problem for the oscillator, will find solutions of the equation for
all values of the variable and parameters. In Section 3 we will consider the
quantum problem for Coulomb-like system. The results will be compared in
section 4, where we will show the one-to one correspondense of the spectra
and proper functions of the Hamiltonians of both problems.

\section{Quantum one-dimentional oscillator}

\label{Osc}

We consider an equation%
\begin{equation}
\partial _{u}^{2}\psi (u)+(W-\lambda u^{2})\psi (u)=0,  \label{Osc1.1}
\end{equation}%
where $\hbar ^{2}W/2\mu $ is complex energy, $\hbar ^{2}\lambda /2\mu $ is a
coupling constant,%
\begin{equation*}
W=|W|e^{i\varphi _{W}},\;+0\leq \varphi _{W}\leq \pi -0,\;\func{Im}W\geq 0,
\end{equation*}

It is convenient to write $\lambda =\varkappa ^{4}$, where%
\begin{equation*}
\varkappa =\left\{ 
\begin{array}{l}
\lambda ^{1/4},\;\varkappa ^{2}=\lambda ^{1/2},\;\lambda \geq 0 \\ 
e^{-i\pi /4}|\lambda |^{1/4},\;\varkappa ^{2}=e^{-i\pi /2}|\lambda
|^{1/2},\;\lambda <0%
\end{array}%
\right. .
\end{equation*}

\subsection{Solutions on the semiaxis $u>0$}

To find the solutions on n the semiaxis $u>0$, we will introduce a new
variable $\rho=(\varkappa u)^{2}$, $\partial_{u}=2\varkappa\sqrt{\rho}%
\partial_{\rho}$, $\partial_{u}^{2}=4[\rho\partial_{\rho}^{2}+(1/2)%
\partial_{\rho}]$, and new function $\phi(\rho)=e^{\rho/2}\psi(u)$. Then we
obtain

\begin{equation}
\rho\partial_{\rho}^{2}\phi(\rho)+(1/2-\rho)\partial_{\rho}\phi(\rho
)-(1/4-w)\phi(\rho)=0,\;w\equiv w_{O}=W/4\varkappa^{2}.  \label{Osc1.1.1}
\end{equation}

Eq. (\ref{Osc1.1.1}) is the equation for confluent hypergeometric functions
with solutions $\Phi(\alpha,\gamma;\rho)$, $\Psi(\alpha,\gamma;\rho),$ in
the terms of which we can express solutions of eq. (\ref{Osc1.1}). We will
use the following solutions:%
\begin{align*}
& O_{+1}(u;W)=\frac{1}{\varkappa}\sqrt{\rho}e^{-\rho/2}\Phi(\alpha
+1/2,3/2;\rho), \\
& O_{+2}(u;W)=e^{-\rho/2}\Phi(\alpha,1/2;\rho), \\
& O_{+3}(u;W)=\pi^{-1/2}\Gamma(\alpha+1/2)e^{-\rho/2}\Psi(\alpha,1/2;\rho)=
\\
& \,=O_{+2}(u;W)-\frac{2\varkappa\Gamma(\alpha+1/2)}{\Gamma(\alpha)}%
O_{+1}(u;W),\;\alpha+1/2\neq-n,n\in\mathbb{Z}_{+}, \\
& \alpha\equiv\alpha_{O}=1/4-w.
\end{align*}

In this section, we will omit the subscript ``$O$''\ meaning, for example, $%
\alpha \equiv \alpha _{O}$, $w\equiv w_{O}$, and so on.

\subsubsection{ Asymptotics}

For $u\rightarrow 0$\ we get 
\begin{align*}
O_{+1}(u;W)& =u+O(u^{3}),\;O_{+2}(u;W)=1+O(u^{2}), \\
O_{+3}(u;W)& =1-\frac{2\varkappa \Gamma (\alpha +1/2)}{\Gamma (\alpha )}%
u+O(u^{2}),\;\alpha +1/2\neq -n,\;n\in \mathbb{Z}_{+}.
\end{align*}

The asymptotics for $u\rightarrow\infty$ and different values of the
parameters we get

$\lambda >0$, $\func{Im}w>0$ or $w=0$ 
\begin{align*}
O_{+1}(u;W)& =\frac{\sqrt{\pi }}{2\varkappa \Gamma (\alpha +1/2)}\rho
^{-1/4-w}e^{\rho /2}(1+O(u^{-2})), \\
O_{+2}(u;W)& =\frac{\sqrt{\pi }}{\Gamma (\alpha )}\rho ^{-1/4-w}e^{\rho
/2}(1+O(u^{-2})), \\
O_{+3}(u;W)& =\frac{\Gamma (\alpha +1/2)}{\sqrt{\pi }}\rho ^{-1/4+w}e^{-\rho
/2}(1+O(u^{-2})),
\end{align*}

$\lambda <0$, $\func{Im}W>0$ or $W=0$%
\begin{align*}
& O_{+1}(u;W)=O(u^{-1/2+\func{Im}W/2\sqrt{|\lambda |}}),%
\;O_{+2}(u;W)=O(u^{-1/2+\func{Im}W/2\sqrt{|\lambda |}}), \\
& O_{+3}(u;W)=O(u^{-1/2-\func{Im}W/2\sqrt{|\lambda |}}).
\end{align*}

The asymptotics for $\lambda =0$ can be obtained as a limit $\lambda
\rightarrow 0$ of corresponding formualae or from explicit expressions for
solutions as $\lambda=0$, 
\begin{align*}
& \left\{ O_{+1}(u;W)\right. _{\lambda =0}=u\Phi (-W/4\varkappa
^{2},3/2;\varkappa ^{2}u^{2})=\frac{1}{\sqrt{W}}\sin (\sqrt{W}u), \\
& \left\{ O_{+2}(u;W)\right. _{\lambda =0}=\Phi (-W/4\varkappa
^{2},1/2;\varkappa ^{2}u^{2})=\cos (\sqrt{W}u), \\
& \left\{ O_{+3}(u;W)\right. _{\lambda =0}=\cos (\sqrt{W}u)-\left. \frac{%
2\varkappa \Gamma (\alpha +1/2)}{\Gamma (\alpha )}\right| _{\varkappa
\rightarrow 0}\frac{1}{\sqrt{W}}\sin (\sqrt{W}u)=e^{i\sqrt{W}u}
\end{align*}%
(where we used a relation $\left. 2\varkappa \Gamma (\alpha +1/2)/\Gamma
(\alpha )\right| _{\varkappa \rightarrow 0}=\sqrt{-W}=-i\sqrt{W}$), which
are in agreement with direct solution of eq. (\ref{Osc1.1}) for $\lambda =0.$

Note, that all solutions of eq. (\ref{Osc1.1}) are square-integrable at the
origin and only solution, $O_{+3}(u;W)$, is square-integrable at infinity
for $\func{Im}W>0$, i. e., $O_{+3}(u;W)\in L^{2}(\mathbb{R}_{+})$ for $\func{%
Im}W>0$.

The functions $O_{+1}$ and $O_{+2}$ are entire functions of $W$ ( for fixed
rest parameters and $u$). They are real for real $W$ and nonnegative $%
\lambda $. If $\lambda $ is negative, then $\varkappa ^{2}$ is pure
imaginary and changes sign under complex conjugation. But the functions $%
O_{+1}$ and $O_{+2}$ are even functions of $\varkappa ^{2}$, that follows
from the relation 9.212.1 of \cite{Grad-Ryzh}%
\begin{align*}
& e^{-\varrho /2}\Phi (\alpha _{O},1/2;z)=e^{-\varkappa ^{2}u^{2}/2}\Phi
(1/4-W/4\varkappa ^{2},1/2;\varkappa ^{2}u^{2})= \\
& \,=e^{\rho /2}\Phi (1/2-\alpha ,1/2;-\rho )=e^{\varkappa ^{2}u^{2}/2}\Phi
(1/4+W/4\varkappa ^{2},1/2;-\varkappa ^{2}u^{2}), \\
& e^{-\rho /2}\Phi (\alpha +1/2,3/2;\rho )=e^{-\varkappa ^{2}u^{2}/2}\Phi
(3/4-W/4\varkappa ^{2},3/2;\varkappa ^{2}u^{2})= \\
& \,=e^{\rho /2}\Phi (3/2-\alpha -1/2,3/2;-\rho )=e^{\varkappa
^{2}u^{2}/2}\Phi (3/4+W/4\varkappa ^{2},3/2;-\varkappa ^{2}u^{2}).
\end{align*}%
Thus, we find that the functions $O_{+1}$ and $O_{+2}$ are real-entire in $W$
for all $\lambda $.

Finally, using the asymptotics for $u\rightarrow 0$ we find the Wronskians
of the solutions%
\begin{align*}
\mathrm{Wr}(O_{+1},O_{+2})& =\mathrm{Wr}(O_{+1},O_{+3})=-1, \\
\mathrm{Wr}(O_{+2},O_{+3})& =-2\varkappa \frac{\Gamma (\alpha +1/2)}{\Gamma
(\alpha )}.
\end{align*}

\subsubsection{Solution on the semiaxis $u<0$}

For $u<0$, we will use the solutions $O_{-k}(u;W)$,%
\begin{equation*}
O_{-k}(u;W)=O_{+k}(|u|;W),\;k=1,2,3,\;u<0.
\end{equation*}

\subsection{Symmetrical operator $\hat{H}_{O}$}

For given a differential operation $\check{H}_{O}$ ($\check{H}$ in what
follows in this section),%
\begin{equation}
\check{H}=-\partial_{u}^{2}+\lambda u^{2},  \label{Osc1.2.1}
\end{equation}
we determine the following symmetrical operator $\hat{H}_{O}\equiv\hat{H}$,

\begin{equation*}
\hat{H}:\left\{ 
\begin{array}{l}
D_{H}=\mathcal{D}(\mathbb{R}\backslash \{0\}), \\ 
\hat{H}\psi (u)=\check{H}\psi (u),\;\forall \psi \in D_{H},%
\end{array}%
\right.
\end{equation*}%
where $\mathcal{D}(\Delta )$ is a space of smooth functions with a compact
support (i.e. which are equal to zero in some neighbourhoods of the
endpoints of the interval $\Delta $).

\subsection{Adjoint operator $\hat{H}_{O}^{+}=\hat{H}_{O}^{\ast }$}

The adjoint operator $\ \hat{H}_{O}^{+}~\ $is

\begin{equation*}
\hat{H}_{O}^{+}\equiv \hat{H}^{+}:\left\{ 
\begin{array}{l}
D_{H^{+}}=\{\psi _{\ast },\psi _{\ast }^{\prime }\;\mathrm{are\;a.c.\;on}%
\mathcal{\;}\mathbb{R}\backslash \{0\},\;\psi _{\ast },\hat{H}^{+}\psi
_{\ast }\in L^{2}(\mathbb{R})\} \\ 
\hat{H}^{+}\psi _{\ast }(u)=\check{H}\psi _{\ast }(u),\;u\in \mathbb{R}%
\backslash \{0\},\;\forall \psi _{\ast }\in D_{H^{+}}%
\end{array}%
\right. .
\end{equation*}%
where a.c. means absolutely continuous.

\subsubsection{Asymptotics of $\protect\psi _{\ast }\in D_{H^{+}}$}

\label{Oasym}

I) $|u|\rightarrow\infty$

Because $V(u)=\lambda u^{2}>-(|\lambda |+1)u^{2}$, we have: $[\psi _{\ast
},\chi _{\ast }](u)\rightarrow 0$ as $u\rightarrow \pm \infty $, $\forall
\psi _{\ast },\chi _{\ast }\in D_{H^{+}}$ \cite{Naima}.

Here $\left[ \chi _{\ast },\psi _{\ast }\right] (u)=\overline{\chi _{\ast
}^{\prime }(u)}\psi _{\ast }(u)-\overline{\chi _{\ast }(u)}\psi _{\ast
}^{\prime }(u).$ .

II) $u\rightarrow+0$

Because $\check{H}\psi _{\ast }\in L^{2}(\mathbb{R})$, we have%
\begin{equation*}
\check{H}\psi _{\ast }(u)=(-\partial _{u}^{2}+\lambda u^{2})\psi _{\ast
}(u)=\eta (u),\;\eta \in L^{2}(\mathbb{R}).
\end{equation*}%
General solution of this equation can be represented in the form%
\begin{align*}
& \psi _{\ast }(u)=a_{+1}O_{+1}(u;0)+a_{+2}O_{+2}(u;0)+I(u), \\
& \psi _{\ast }^{\prime }(u)=a_{+1}O_{+1}^{\prime
}(u;0)+a_{+2}O_{+2}^{\prime }(u;0)+I^{\prime }(u),
\end{align*}%
where

\begin{align*}
& I(u)=O_{+2}(u;0)\int_{0}^{u}O_{+1}(v;0)\eta
(v)dv-O_{+1}(u;0)\int_{0}^{u}O_{+2}(v;0)\eta (v)dv, \\
& I^{\prime }(u)=O_{+2}^{\prime }(u;0)\int_{0}^{u}O_{+1}(v;0)\eta
(v)dv-O_{+1}^{\prime }(u;0)\int_{0}^{u}O_{+2}(v;0)\eta (v)dv.
\end{align*}%
We obtain with the help of the Cauchy-Bunyakovskii inequality:%
\begin{equation*}
I(u)=O(u^{3/2}),\;I^{\prime }(u)=O(u^{1/2}),\;u\rightarrow +0,
\end{equation*}%
such that we find%
\begin{align*}
& \psi _{\ast }(u)=a_{+1}u+a_{+2}+O(u^{3/2}),\;\psi _{\ast }^{\prime
}(u)=a_{+1}+O(u^{1/2}),\;u\rightarrow +0, \\
& a_{+2}=\psi _{\ast }(+0),\;a_{+1}=\psi _{\ast }^{\prime }(+0).
\end{align*}

III) $u\rightarrow-0$ Analogously, we obtain for $u\rightarrow-0$:%
\begin{align*}
& \psi_{\ast}(u)=-a_{-1}u+a_{-2}+O(|u|^{3/2}),\;\psi_{\ast}^{\prime
}(u)=-a_{-1}+O(|u|^{1/2}),\;u\rightarrow-0, \\
& a_{-2}=\psi_{\ast}(-0),\;a_{-1}=-\psi_{\ast}^{\prime}(-0).
\end{align*}

\subsection{Sesquilinear form}

Sesquilinear form of adjoint operator $\hat{H}^{+},$ $\omega _{H^{+}}(\psi
_{\ast },\chi _{\ast })$ is defined as%
\begin{align*}
& \omega _{H^{+}}(\chi _{\ast },\psi _{\ast })=\omega _{+H^{+}}(\chi _{\ast
},\psi _{\ast })+\omega _{-H^{+}}(\chi _{\ast },\psi _{\ast }), \\
& \omega _{+H^{+}}(\chi _{\ast },\psi _{\ast })=\int_{0}^{\infty }\left[ 
\overline{\chi _{\ast }(u)}\check{H}\psi _{\ast }(u)-\overline{\check{H}\chi
_{\ast }(u)}\psi _{\ast }(u)\right] du= \\
& \,=-\left. \left[ \chi _{\ast },\psi _{\ast }\right] (u)\right|
_{u\rightarrow +0}=\overline{a_{\chi _{\ast }+2}}a_{\psi _{\ast }+1}-%
\overline{a_{\chi _{\ast }+1}}a_{\psi _{\ast }+2}, \\
& \omega _{-H^{+}}(\chi _{\ast },\psi _{\ast })=\int_{-\infty }^{0}\left[ 
\overline{\chi _{\ast }(u)}\check{H}\psi _{\ast }(u)-\overline{\check{H}\chi
_{\ast }(u)}\psi _{\ast }(u)\right] du= \\
& \,=\left. \left[ \chi _{\ast },\psi _{\ast }\right] (u)\right|
_{u\rightarrow -0}=\overline{a_{\chi _{\ast }-2}}a_{\psi _{\ast }-1}-%
\overline{a_{\chi _{\ast }-1}}a_{\psi _{\ast }-2}, \\
& \left[ \chi _{\ast },\psi _{\ast }\right] (u)=\overline{\chi _{\ast
}^{\prime }(u)}\psi _{\ast }(u)-\overline{\chi _{\ast }(u)}\psi _{\ast
}^{\prime }(u).
\end{align*}%
Thus we have%
\begin{align*}
& \omega _{H_{O}^{+}}(\chi _{\ast },\psi _{\ast })=\overline{\mathbf{a}%
_{\chi _{\ast }2}}\mathbf{a}_{\psi _{\ast }1}-\overline{\mathbf{a}_{\chi
_{\ast }1}}\mathbf{a}_{\psi _{\ast }2}=\frac{i}{2\kappa _{0}}\left( 
\overline{\mathbf{b}_{\chi _{\ast }}}\mathbf{b}_{\psi _{\ast }}-\overline{%
\mathbf{d}_{\chi _{\ast }}}\mathbf{d}_{\psi _{\ast }}\right) , \\
& \mathbf{a}_{1}=\left( 
\begin{array}{c}
a_{+1} \\ 
a_{-1}%
\end{array}%
\right) ,\;\mathbf{a}_{2}=\left( 
\begin{array}{c}
a_{+2} \\ 
a_{-2}%
\end{array}%
\right) , \\
& \mathbf{b}=\left( 
\begin{array}{c}
b_{+} \\ 
b_{-}%
\end{array}%
\right) =\mathbf{a}_{1}+i\kappa _{0}\mathbf{a}_{2},\;\mathbf{d}=\left( 
\begin{array}{c}
d_{+} \\ 
d_{-}%
\end{array}%
\right) =\mathbf{a}_{1}-i\kappa _{0}\mathbf{a}_{2},
\end{align*}%
where $\kappa _{0}$ is arbitrary, but fixed parameter of dimensionality of
inverse length introduced by dimensional reasons.

\subsection{Self-adjoint hamiltonians}

Because all self-adjoint (s.a.) hamiltonians, $\hat{H}_{O\mathfrak{e}}$ ($%
\equiv\hat{H}_{\mathfrak{e}}$), act on their domains as $\check{H}$, we
should specify definition domains only. The definition domain $D_{H_{%
\mathfrak{e}}}$ of s.a. operator $\hat{H}_{\mathfrak{e}}$ is determined by
condition%
\begin{equation*}
\omega_{H^{+}}(\chi,\psi)=0,\;\forall\chi,\psi\in D_{H_{\mathfrak{e}}},
\end{equation*}
from which it follows%
\begin{equation}
\mathbf{d}_{\psi}=U\mathbf{b}_{\psi},\;\forall\psi\in D_{H_{\mathfrak{e}}},
\label{Osc1.5.1}
\end{equation}
where $U$ is an arbitrary, but fixed for given extension, unitary $(2\times
2)$-matrix, $U^{+}U=1$. Thus, any s.a. hamiltonian is determined by
assignment of unitary matrix $U$ (we will denote the corresponding s.a.
hamiltonian by $\hat{H}_{OU}\equiv\hat{H}_{U}$),%
\begin{equation*}
\hat{H}_{U}^{+}:\left\{ 
\begin{array}{l}
D_{H_{U}}\equiv D_{U}=\{\psi:\;\psi\in D_{\check{H}}^{\ast},\;\mathbf{d}%
_{\psi}=U\mathbf{b}_{\psi}\} \\ 
\hat{H}_{U}\psi(u)=\check{H}\psi(u),\;u\in\mathbb{R}\backslash\{0\},\;%
\forall \psi\in D_{U}%
\end{array}
\right. .
\end{equation*}
Thus, there exists a $U(2)$-family of s.a. extensions of the initial
symmetric operator $\hat{H}$.

\subsection{Parity conserving extensions}

We will further restrict ourselves to the s.a. extensions conserving parity, 
$[\hat{P},\hat{H}_{{\large U}}]=0$, where.$\hat{P}$ is the parity operator
that acts on functions $\psi \left( x\right) $ in $L^{2}(\mathbb{R})$ as%
\begin{equation}
\hat{P}\psi \left( u\right) =\psi \left( -u\right) .  \label{Osc1.6.1}
\end{equation}%
The Hilbert space $L^{2}(\mathbb{R})$ can be decomposed in the direct
orthogonal sum of a subspace $L_{s}^{2}(\mathbb{R})$ symmetric functions and
a subspace $L_{a}^{2}(\mathbb{R})$ of antisymmetric functions, such that $%
L^{2}(\mathbb{R})=L_{s}^{2}(\mathbb{R})\oplus L_{a}^{2}(\mathbb{R}),$ 
\begin{align*}
& \psi \in L^{2}(\mathbb{R}),\ \ \psi =\psi _{s}+\psi _{a}\ ,\ \ \psi
_{s}\in L_{s}^{2}(\mathbb{R}),\ \ \psi _{a}\in L_{a}^{2}(\mathbb{R}), \\
& \hat{P}\psi _{s}=\psi _{s},\ \ \hat{P}\psi _{a}=-\psi _{a}\ .
\end{align*}

One can easily see that operators $\hat{H}_{O}$ and $\hat{H}_{O}^{+}$
commute with $\hat{P},$%
\begin{equation*}
\lbrack \hat{P},\hat{H}]=[\hat{P},\hat{H}^{+}]=0.
\end{equation*}%
This means that the operators $\hat{H}$ and $\hat{H}^{+}$ can be represented
in the form of direct sum of their parts in the corresponding subdomains of
symmetric and antisymmetric functions:%
\begin{align*}
& \hat{H}=\hat{H}_{s}\oplus \hat{H}_{a},\;D_{H}=D_{Hs}\oplus
D_{Ha},\;D_{Hs,a}=\mathcal{D}_{s,a}\left( \mathbb{R}\backslash \{0\}\right) ,
\\
& \hat{H}\psi =\hat{H}_{s}\psi _{s}+\hat{H}_{a}\psi _{a},\;\psi =\psi
_{s}+\psi _{a},\ \psi \in D_{H},\ \psi _{s,a}\in D_{Hs,a},
\end{align*}%
where $\mathcal{D}_{s,a}\left( \mathbb{R}\backslash \{0\}\right) $ are
subspaces of symmetric (antisymmetric) functions in $\mathcal{D}\left( 
\mathbb{R}\backslash \{0\}\right) .$ Similar decompositions hold true for
the adjoint operator $\hat{H}^{+}$,%
\begin{equation*}
\ \hat{H}^{+}=\hat{H}_{s}^{+}\oplus \hat{H}_{a}^{+}\ ,\ D_{H^{+}}\left( 
\mathbb{R}\backslash \{0\}\right) =D_{H^{+}}\left( \mathbb{R}\backslash
\{0\}\right) _{s}\oplus D_{H^{+}}\left( \mathbb{R}\backslash \{0\}\right)
_{a}\ ,
\end{equation*}%
where $D_{H^{+}}\left( \mathbb{R}\backslash \{0\}\right) _{s,a}$ are
subspaces of symmetric (antisymmetric) functions in $D_{H^{+}}\left( \mathbb{%
R}\backslash \{0\}\right) $:%
\begin{equation*}
D_{H^{+}}\left( \mathbb{R}\backslash \{0\}\right) _{s,a}=\left\{ \psi _{\ast
s,a}:\psi _{\ast s,a},\psi _{\ast s,a}^{\prime }\ \mathrm{are\ a.c.\ on}\ 
\mathbb{R}\backslash \{0\},\ \psi _{\ast s,a},\hat{H}_{s,a}^{+}\psi _{\ast
s,a}\in L_{s,a}^{2}(\mathbb{R})\right\}
\end{equation*}

Because the operator $\hat{P}$ is bounded, $||\hat{P}||=1$, and $\hat{P}%
^{2}=1$, the assertion that $\hat{P}$ commutes with $\hat{H}_{{\large U}}$
means 
\begin{equation*}
\lbrack \hat{P},\hat{H}_{{\large U}}]=0\Longrightarrow \hat{H}_{U}=\hat{H}%
_{sU}\oplus \hat{H}_{aU}\ ,
\end{equation*}%
where operators $\hat{H}_{s,aU}$ are s.a. extensions of the operators $%
\hat{H}_{s,a}$ . In turn, if $\hat{H}_{s,aU}$ are s.a. extensions of $\hat{H}%
_{s,a} $ in $L_{s,a}^{2}(\mathbb{R})$, then the operator $\hat{H}_{U}=\hat{H}%
_{sU}\oplus \hat{H}_{aU}$ is a s.a. extension of $\hat{H}$ in $L^{2}(\mathbb{%
R})$ which commutes with $\hat{P}$. Thus, it is enough to describe all s.a.
extensions of operators $\hat{H}_{s,a}$ in the subspaces $L_{s,a}^{2}(%
\mathbb{R})$ to find all commuting with $\hat{P}$ s.a. extensions $\hat{H}_{%
{\large U}}$ of the operator $\hat{H}$.

First, we find the general form of matrix $U=U_{P}$ conserving (commuting
with) the parity $\hat{P}$.

The condition: $U_{P}$ commutes with $\hat{P}$, means that rel. (\ref%
{Osc1.5.1}) is valid for the functions $\psi _{s,a}\in L_{s,a}^{2}(\mathbb{R}%
)$. The functions $\psi _{s,a}$ have the properties 
\begin{equation}
a_{s,a-2}=\pm a_{s,a+2},\ a_{s,a-1}=\pm a_{s,a+1},  \label{Osc1.6.1a}
\end{equation}%
such that doublets $\mathbf{d}_{\psi _{s,a}}$ and $\mathbf{b}_{\psi _{s,a}}$%
\textbf{\ }have the form,%
\begin{align*}
& \mathbf{b}_{\psi _{s,a}}=\sqrt{2}[a_{s,a+1}+i\kappa _{0}a_{s,a+2}]\mathbf{n%
}_{s,a},\;\mathbf{d}_{\psi _{s,a}}=\sqrt{2}[a_{s,a+1}-i\kappa _{0}a_{s,a+2}]%
\mathbf{n}_{s,a}, \\
& \mathbf{n}_{s}=\left( 
\begin{array}{c}
1/\sqrt{2} \\ 
1/\sqrt{2}%
\end{array}%
\right) ,\ \mathbf{n}_{a}=\left( 
\begin{array}{c}
1/\sqrt{2} \\ 
-1/\sqrt{2}%
\end{array}%
\right) .
\end{align*}%
The condition (\ref{Osc1.5.1}) gives for such doublets%
\begin{align}
& U_{P}\mathbf{n}_{s,a}=\lambda _{s,a}\mathbf{n}_{s,a},\   \label{Osc1.6.2}
\\
& \lambda _{s,a}=\frac{a_{s,a+1}-i\kappa _{0}a_{s,a+2}}{a_{s,a+1}+i\kappa
_{0}a_{s,a+2}}=\frac{\psi _{s,a}^{\prime }(+0)-i\kappa _{0}\psi _{s,a}(+0)}{%
\psi _{s,a}^{\prime }(+0)+i\kappa _{0}\psi _{s,a}(+0)}=e^{i\varphi _{s,a}},\
0\leq \varphi _{s,a}\leq 2\pi ,  \notag
\end{align}%
i. e., orthonormalized vectors $\mathbf{n}_{s,a}$ must be eigenvectors of
matrix $U$. General form of matrices $U=U_{P}$ satisfying condition (\ref%
{Osc1.6.2}) is 
\begin{equation}
U_{P}=\lambda _{s}\mathbf{n}_{s}\otimes \mathbf{n}_{s}+\lambda _{a}\mathbf{n}%
_{a}\otimes \mathbf{n}_{a}.  \label{Osc1.6.3}
\end{equation}%
The inverse statement is true as well. Namely, if matrix $U$ has the form (%
\ref{Osc1.6.3}) then the subspaces $L_{s,a}^{2}(\mathbb{R})$ reduce the
corresponding s.a. hamiltonian $\hat{H}_{U_{P}}$, i. e., the hamiltonian $%
\hat{H}_{U_{P}}$ commutes with parity operator $\hat{P}$.

In the terms of the asymptotical boundary (a.b.) conditions, such a form of
the matrix $U_{P}$ means the following:%
\begin{equation}
a_{s,a+1}\cos\zeta_{s,a}=\kappa_{0}a_{s,a+2}\sin\zeta_{s,a},\;|\zeta
_{s,a}|\leq\pi/2,\ \zeta_{s,a}=-\pi/2\sim\zeta_{s,a}=\pi/2,
\label{Osc1.6.4a}
\end{equation}
or%
\begin{equation}
\psi_{s,a}(u)=\left\{ 
\begin{array}{l}
a(\kappa_{0}u\sin\zeta_{s,a}+\cos\zeta_{s,a})+O(u^{3/2}),\;u>0 \\ 
\pm a(\kappa_{0}|u|\sin\zeta_{s,a}+\cos\zeta_{s,a})+O(u^{3/2}),\;u<0%
\end{array}
\right. ,\;u\rightarrow0,  \label{Osc1.6.4b}
\end{equation}
where $\zeta_{s,a}=\varphi_{s,a}/2-\pi/2$. The inverse statement is true as
well. Namely, if matrix $U$ gives the boundary condition of the form (\ref%
{Osc1.6.4b}) (or (\ref{Osc1.6.4a})) then that matrix $U$ has the form (\ref%
{Osc1.6.3}) with $\varphi_{s,a}=2\zeta_{s,a}+\pi$. In what follows, we
change the notation of s.a. operator $\hat{H}_{U_{P}}$ for $\hat{H}%
_{\zeta_{s,a}}$.

\subsection{Extensions on semiaxis $\mathbb{R}_{+}$}

To extend the adjoint operator on semiaxis $\mathbb{R}_{+}$\ \ define for
the differential operation $\check{h}_{O}$ ($\equiv\check{h}$)

\begin{equation*}
\check{h}=\check{H}=-\partial _{u}^{2}+\lambda u^{2},
\end{equation*}%
a symmetrical operator $\hat{h}_{O}$ ($\equiv \hat{h}$)

\begin{equation*}
\hat{h}:\left\{ 
\begin{array}{l}
D_{h}=\mathcal{D}(\mathbb{R}_{+}) \\ 
\hat{h}\psi (u)=\check{h}\psi (u),\;\forall \psi \in D_{h},%
\end{array}%
\right.
\end{equation*}%
and the adjoint operator $\hat{h}_{O}^{+}$

\begin{equation*}
\hat{h}_{O}^{+}\equiv \hat{h}^{+}:\left\{ 
\begin{array}{l}
D_{h^{+}}=\{\psi _{\ast },\psi _{\ast }^{\prime }\;\mathrm{are\;a.c.\;on}%
\mathcal{\;}\mathbb{R}_{+},\;\psi _{\ast },\check{H}_{O}\psi _{\ast }\in
L^{2}(\mathbb{R}_{+})\} \\ 
\hat{h}^{+}\psi _{\ast }(u)=\check{h}\psi _{\ast }(u),\;\forall \psi _{\ast
}\in D_{h^{+}}%
\end{array}%
\right. .
\end{equation*}

Literally repeating the considerations of subsec.\ref{Oasym} we obtain the
asymptotics:

I) $u\rightarrow\infty$

$[\psi _{\ast },\chi _{\ast }](u)\rightarrow 0$, $\forall \psi _{\ast },\chi
_{\ast }\in D_{h^{+}}$.

II) $u\rightarrow 0$%
\begin{align*}
& \psi _{\ast }(u)=a_{1}u+a_{2}+O(u^{3/2}),\;\psi _{\ast }^{\prime
}(u)=a_{1}+O(u^{1/2}), \\
& a_{2}=\psi _{\ast }(0),\;a_{1}=\psi _{\ast }^{\prime }(0).
\end{align*}%
For the sesquilinear form $\omega _{h^{+}}(\psi _{\ast },\chi _{\ast })$ we
get%
\begin{align*}
& \omega _{h^{+}}(\chi _{\ast },\psi _{\ast })=\int_{0}^{\infty }\left[ 
\overline{\chi _{\ast }(u)}\check{h}\psi _{\ast }(u)-\overline{\check{h}\chi
_{\ast }(u)}\psi _{\ast }(u)\right] du= \\
& \,=-\left. \left[ \chi _{\ast },\psi _{\ast }\right] (u)\right|
_{u\rightarrow 0}=\overline{a_{\chi _{\ast }2}}a_{\psi _{\ast }1}-\overline{%
a_{\chi _{\ast }1}}a_{\psi _{\ast }2}= \\
& =\frac{i}{2\kappa _{0}}\left( \overline{b_{\chi _{\ast }}}b_{\psi _{\ast
}}-\overline{d_{\chi _{\ast }}}d_{\psi _{\ast }}\right) ,\;b=a_{1}+i\kappa
_{0}a_{2},\;d=a_{1}-i\kappa _{0}a_{2}.
\end{align*}

\subsubsection{Self-adjoint hamiltonians}

Because all s.a. hamiltonians, $\hat{h}_{O\mathfrak{e}}$ ($\equiv \hat{h}_{%
\mathfrak{e}}$), act on its domains as $\check{h}$, we should specify
definition domains only. The definition domain $D_{h_{\mathfrak{e}}}$ of
s.a. operator $\hat{h}_{\mathfrak{e}}$ is determined by condition%
\begin{equation*}
\omega _{h^{+}}(\chi ,\psi )=0,\;\forall \chi ,\psi \in D_{h_{\mathfrak{e}}},
\end{equation*}%
from which it follows%
\begin{equation*}
d_{\psi }=e^{i\varphi }b_{\psi },\;\forall \psi \in D_{h_{\mathfrak{e}%
}},\;0\leq \varphi \leq 2\pi ,\;0\sim 2\pi ,
\end{equation*}%
or, equivalent%
\begin{equation*}
a_{1}\cos \zeta =\kappa _{0}a_{2}\sin \zeta ,\;\zeta \in S(-\pi /2,\pi
/2),\;\zeta =\varphi /2-\pi /2.
\end{equation*}%
Thus, any s.a. hamiltonian is determined by assignment of unitary matrix $%
U(1)=e^{i\varphi }$ (we will denote the corresponding s.a. hamiltonian by $%
\hat{h}_{O\zeta }$),%
\begin{equation*}
\hat{h}_{O\zeta }\equiv \hat{h}_{\zeta }:\left\{ 
\begin{array}{l}
D_{h_{\zeta }}\equiv D_{\zeta }=\{\psi :\;\psi \in D_{\check{h}}^{\ast
},\;a_{1}\cos \zeta =\kappa _{0}a_{2}\sin \zeta \} \\ 
\hat{h}_{\zeta }\psi (u)=\check{h}\psi (u),\;\forall \psi \in D_{\zeta }%
\end{array}%
\right. .
\end{equation*}%
Equivalently, the boundary condition for $\psi \in D_{\zeta }$ can be
represented in the form

\begin{equation}
\psi (u)=a(\kappa _{0}u\sin \zeta +\cos \zeta )+O(u^{3/2}),\;u\rightarrow 0.
\label{Osc1.7.5.1}
\end{equation}

Thus, there exists a $U(1)$-family of s.a. extensions $\hat{h}_{\zeta }$ of
the initial symmetric operator $\hat{h}$.

\subsection{Self-adjoint extensions of $\hat{H}_{s}$}

The Hilbert space $L_{s}^{2}(\mathbb{R})$ is the space of all symmetric
functions that are square integrable on $\mathbb{R}$. These functions obey
the relations 
\begin{equation*}
\psi(+0)=\psi(-0)=\psi(0),\ \psi^{\prime}(+0)=-\psi^{\prime}(-0),\ \forall
\psi\in L_{s}^{2}(\mathbb{R})\ ,
\end{equation*}
(see (\ref{Osc1.6.1a})) which implies%
\begin{equation}
(\chi,\psi)=2(\chi,\psi)_{+}\ ,\
\omega_{H^{+}}(\chi,\psi)=2\omega_{H^{+}}(\chi,\psi)_{+}=2\omega_{h^{+}}(%
\chi,\psi),  \label{Osc1.8.1}
\end{equation}

where%
\begin{equation}
(\chi,\psi)_{+}=\int_{0}^{\infty}\overline{\chi(u)}\psi\left( u\right) du,
\label{Osc1.8.2}
\end{equation}
and $\omega_{H^{+}}(\chi,\psi)_{+}=\omega_{h^{+}}(\chi,\psi)$ is the
sesquilinear form with respect to the scalar product (\ref{Osc1.8.2}).

Let us consider the isometry $T$: $\psi \in \mathbb{R}\overset{T}{%
\longrightarrow }\sqrt{2}\psi $, $\psi \in \mathbb{R}_{+}$. Then%
\begin{equation}
D_{H_{s}}\overset{T}{\longrightarrow }D_{h}=\mathcal{D}(\mathbb{R}_{+}),\ \
D_{H_{s}^{+}}=D_{\check{H}}^{\ast }\left( \mathbb{R}\backslash \{0\}\right) 
\overset{T}{\longrightarrow }D_{h^{+}}.  \label{Osc1.8.3}
\end{equation}

It follows from eqs. (\ref{Osc1.8.1}) and (\ref{Osc1.8.3}) that there is
one-to-one correspondence (the isometry $T$) between s.a. extensions $\hat{H}%
_{\zeta_{s}}$ of the symmetric operator $\hat{H}_{s}$ in $L_{s}^{2}(\mathbb{R%
})$ and s.a. extensions $\hat{h}_{\zeta}$ of the symmetric operator $\hat{h}$
in $L^{2}(\mathbb{R}_{+})$: $\hat{H}_{\zeta_{s}}\overset{T}{%
\Longleftrightarrow}\hat{h}_{\zeta}$, $\zeta_{s}=\zeta$. Thus, the spectral
analysis of s.a. operator $\hat{H}_{\zeta_{s}}$ in $L_{s}^{2}(\mathbb{R})$
is reduced to the spectral analysis of s.a. operator $\hat{h}_{\zeta}$, $%
\zeta_{s}=\zeta$, in $L^{2}(\mathbb{R}_{+})$. Below, we represent this
analysis.

\subsubsection{Guiding functional}

To perform the analysis we have to define the guiding functional $\Phi
_{O\zeta }(\xi ;W)\equiv \Phi _{\zeta }(\xi ;W)$ \cite{Naima,AkhGlaz} 
\begin{align*}
& \Phi _{\zeta }(\xi ;W)=\int_{0}^{\infty }U_{\zeta }(u;W)\xi (u)du,\;\xi
\in \mathbb{D}_{\zeta }=D_{r}(\mathbb{R}_{+})\cap D_{h_{\zeta }}, \\
& U_{\zeta }(u;W)\equiv U_{O\zeta }(u;W)=\kappa _{0}O_{+1}(u;W)\sin \zeta
+O_{+2}(u;W)\cos \zeta , \\
& D_{r}(a,b)=\{\psi (u):\;\mathrm{supp}\psi \subseteq \lbrack a,\beta _{\psi
}],\;\beta _{\psi }<b.
\end{align*}%
Note that $U_{\zeta }(u;W)$ is real-entire solution of eq. (\ref{Osc1.1})
and satisfies the boundary conditions (\ref{Osc1.7.5.1}).

The guiding functional $\Phi _{\zeta }(\xi ;W)$ satisfies the properties 1)-
3) of \cite{Naima,AkhGlaz} . We will call the guiding functional $\Phi
_{\zeta }(\xi ;W)$ with those properties ''simple''. It follows \cite%
{Naima,AkhGlaz} that the spectrum of $\hat{h}_{\zeta }$ is simple.

\subsubsection{Green function $G_{O\protect\zeta }(u,v;W)$, spectral
function $\protect\sigma _{O\protect\zeta }(E)$}

The Green function $G_{O\zeta }(u,v;W)\equiv G_{\zeta }(u,v;W)$ is the
kernel of the integral representation 
\begin{equation*}
\psi (u)=\int_{0}^{\infty }G_{\zeta }(u,v;W)\eta (v)dv,\;\eta \in L^{2}(%
\mathbb{R}_{+}),
\end{equation*}%
of unique solution of an equation%
\begin{equation}
(\hat{h}_{\zeta }^{+}-W)\psi (u)=\eta (u),\;\func{Im}W>0,  \label{Osc1.8.2.1}
\end{equation}%
for $\psi \in D_{\zeta }$, that is, .$\psi \in L^{2}(\mathbb{R}_{+})$ and $%
\psi $ satisfies the boundary conditions (\ref{Osc1.7.5.1}).\ We find%
\begin{align}
& G_{\zeta }(u,v;W)=\Omega _{\zeta }(W)U_{\zeta }(u;W)U_{\zeta }(v;W)-\frac{1%
}{\kappa _{0}}\left\{ 
\begin{array}{c}
\tilde{U}_{\zeta }(u;W)U_{\zeta }(v;W),\;u>v \\ 
U_{\zeta }(u;W)\tilde{U}_{\zeta }(v;W),\;u<v%
\end{array}%
\right. ,  \label{Osc1.8.2.2} \\
& \Omega _{\zeta }(W)\equiv \Omega _{O\zeta }(W)=\frac{\tilde{\omega}_{\zeta
}(W)}{\kappa _{0}\omega _{\zeta }(W)},\;\omega _{O\zeta }(W)\equiv \omega
_{\zeta }(W)=\sin \zeta +\tilde{\gamma}(W)\cos \zeta ,  \notag \\
& \tilde{\omega}_{\zeta }(W)\equiv \tilde{\omega}_{O\zeta }(W)=\cos \zeta -%
\tilde{\gamma}(W)\sin \zeta ,\;\tilde{\gamma}(W)\equiv \tilde{\gamma}_{O}(W)=%
\frac{2\varkappa }{\kappa _{0}}\gamma (\alpha ),  \notag \\
& \gamma (\alpha )\equiv \gamma _{O}(\alpha _{O})=\frac{\Gamma (\alpha +1/2)%
}{\Gamma (\alpha )},  \notag
\end{align}%
where we used relations%
\begin{align*}
& \kappa _{0}O_{+1}(u;W)=U_{\zeta }(u;W)\sin \zeta +\tilde{U}_{\zeta
}(u;W)\cos \zeta , \\
& O_{+2}(u;W)=U_{\zeta }(u;W)\cos \zeta -\tilde{U}_{\zeta }(u;W)\sin \zeta ,
\\
& O_{+3}(u;W)=U_{\zeta }(u;W)\tilde{\omega}_{\zeta }(W)-\tilde{U}_{\zeta
}(u;W)\omega _{\zeta }(W).
\end{align*}%
Note that $\tilde{U}_{\zeta }(u;W)$ is real-entire solution of eq. (\ref%
{Osc1.1}) such that the last term in the r.h.s. of eq. (\ref{Osc1.8.2.2}) is
real for $W=E_{O}\equiv E$ ($\func{Im}W=0$). From the relation \cite%
{Naima,AkhGlaz}

\begin{equation*}
U_{\zeta }^{2}(u_{0};E)\sigma _{\zeta }^{\prime }(E)=\frac{1}{\pi }\func{Im}%
G_{\zeta }(u_{0}-0,u_{0}+0;E+i0),
\end{equation*}%
where $f(E+i0)\equiv \lim_{\varepsilon \rightarrow +0}f(E+i\varepsilon )$, $%
\forall f(W)$, we find%
\begin{equation*}
\sigma _{O\zeta }(E)\equiv \sigma _{\zeta }^{\prime }(E)=\frac{1}{\pi }\func{%
Im}\Omega _{\zeta }(E+i0).
\end{equation*}

Now we proceed to defining the spectrum of the theory.

\subsubsection{Spectrum, $\protect\lambda >0$}

We have $\varkappa=\lambda^{1/4}>0$, $W=E+i\varepsilon$, $w(E)=w|_{W=E}=E/4%
\sqrt {\lambda}$

{\ $\mathbf{a)}\,\zeta =\pi /2$}

\begin{equation*}
\Omega _{\pi /2}(W)=-\frac{\tilde{\gamma}(E+i\varepsilon )}{\kappa _{0}}=-%
\frac{2\lambda ^{1/4}\Gamma (3/4-(E+i\varepsilon )/4\sqrt{\lambda })}{\kappa
_{0}^{2}\Gamma (1/4-(E+i\varepsilon )/4\sqrt{\lambda })}
\end{equation*}%
The function $\Omega _{\pi /2}(E)$ is real for $E$ where $|\Omega _{O\pi
/2}(E)|<\infty $. Therefore, $\func{Im}\Omega _{\pi /2}(E+i0)$ can be not
equal to zero only in the points $\Omega _{\pi /2}(E)=\pm \infty $, i.e., in
the points $\alpha +1/2=-n$, $w(\vartheta _{On})=w_{\pi /2|n}=n+3/4$, $n\in 
\mathbb{Z}_{+}$, or%
\begin{equation*}
\vartheta _{On}=\vartheta _{n}=4\lambda ^{1/2}n+3\lambda ^{1/2}=2\lambda
^{1/2}[(2n+1)+1/2].
\end{equation*}%
In the neighborhood of the points $\vartheta _{n}$ we have ($W=\vartheta
_{n}+\Delta $, $\Delta =E-\vartheta _{n}+i\varepsilon $, $\alpha =-1/2-n-%
\tilde{\Delta},\;\tilde{\Delta}=\Delta /4\sqrt{\lambda }$)%
\begin{align*}
& \func{Im}\Omega _{\pi /2}(E+i0)=-\frac{2\lambda ^{1/4}}{\kappa
_{0}^{2}\Gamma (-1/2-n)}\func{Im}\Gamma (-n-\tilde{\Delta})|_{\varepsilon
\rightarrow +0}= \\
& \,=\pi Q_{\pi /2|n}^{2}\delta (E-\vartheta _{n}),\;Q_{\pi /2|n}=\frac{2}{%
\kappa _{0}}\left[ \frac{\lambda ^{3/4}(2n+1)!!}{\sqrt{\pi }(2n)!!}\right]
^{1/2}.
\end{align*}%
Finally, we find%
\begin{equation*}
\sigma _{\pi /2}^{\prime }(E)=\sum_{n=0}^{\infty }Q_{\pi /2|n}^{2}\delta
(E-\vartheta _{n}),\;\mathrm{spec}\hat{h}_{\pi /2}=\{\vartheta _{n},\;n\in 
\mathbb{Z}_{+}\}.
\end{equation*}%
A complete orthonormalized system of (generalized) eigenfunctions of $\hat{h}%
_{\pi /2}$ is $\{U_{\pi /2|n}(u)=Q_{\zeta |n}k_{0}O_{+1}(u;\vartheta
_{n}),\;n\in \mathbb{Z}_{+}\}$.

We obtain the same results for the case $\zeta =-\pi /2$.

{\ $\mathbf{b)}\,\zeta=0$}

\begin{equation*}
\Omega _{0}(E+i\varepsilon )=\frac{1}{\kappa _{0}\tilde{\gamma}%
(E+i\varepsilon )}=\frac{\Gamma (1/4-(E+i\varepsilon )/4\sqrt{\lambda })}{%
2\lambda ^{1/4}\Gamma (3/4-(E+i\varepsilon )/4\sqrt{\lambda })}.
\end{equation*}%
The function $\Omega _{0}(E)$ is real for $E$ if $|\Omega _{0}(E)|<\infty $.
Therefore, $\func{Im}\Omega _{0}(E+i0)$ can be not equal to zero only in the
point $\Omega _{0}(E)=\pm \infty $, i. e., in the points $\alpha =-n$, $%
w(E_{0|n})=w_{0|n}=n+1/4$, $n\in \mathbb{Z}_{+}$, or%
\begin{equation*}
E_{0|n}=4\lambda ^{1/2}n+\lambda ^{1/2}=2\lambda ^{1/2}(2n+1/2).
\end{equation*}%
In the neighborhood of the points $E_{0|n}$ we have 
\begin{align*}
& \func{Im}\Omega _{O0}(E+i0)=\frac{1}{2\lambda ^{1/4}\Gamma (1/2-n)}\func{Im%
}\Gamma (-n-\tilde{\Delta})|_{\varepsilon \rightarrow +0}= \\
& \,=\pi Q_{0|n}^{2}\delta (E-E_{0|n}),\;Q_{0|n}=\left[ \frac{%
2(2n-1)!!\lambda ^{1/4}}{\sqrt{\pi }(2n)!!}\right] ^{1/2}.
\end{align*}%
Finally, we find%
\begin{equation*}
\sigma _{0}^{\prime }(E)=\sum_{n=0}^{\infty }Q_{0|n}^{2}\delta (E-E_{0|n}),\;%
\mathrm{spec}\hat{h}_{0}=\{E_{0|n,}\;n\in \mathbb{Z}_{+}\}.
\end{equation*}

A complete orthonormalized system of (generalized) eigenfunctions of $\hat{h}%
_{0}$ is $\{U_{0|n}(u)=Q_{0|n}O_{+2}(u;E_{0|n}),\;n\in \mathbb{Z}_{+}\}$.

$\mathbf{c)}$ General case of $|\zeta|<\pi/2$

In this case we have

\begin{equation*}
\sigma _{\zeta }^{\prime }(E)=\frac{1}{\pi \kappa _{0}\cos ^{2}\zeta }\func{%
Im}\frac{1}{\tilde{\gamma}(E+i0)+\tan \zeta }.
\end{equation*}%
The function $\tilde{\gamma}(E)$ is real for real $E$. Therefore, $\sigma
_{\zeta }^{\prime }(E)$ can be not equal to zero only in the points 
\begin{equation}
\tilde{\gamma}(E_{\zeta |n})=-\tan \zeta .  \label{Osc1.8.3.3.1}
\end{equation}

For the derivative of spectral function $\sigma _{\zeta }^{\prime }(E)$ we
find%
\begin{align*}
& \sigma _{\zeta }^{\prime }(E)=\sum_{n=0}^{\infty }Q_{\zeta |n}^{2}\delta
(E-E_{\zeta |n}),\;Q_{\zeta |n}=\left[ -\frac{1}{\kappa _{0}\tilde{\gamma}%
^{\prime }(E_{\zeta |n})\cos ^{2}\zeta }\right] ^{1/2}, \\
& \tilde{\gamma}^{\prime }(E_{\zeta |n})<0,\;\partial _{\zeta }E_{\zeta
|n}=-1/[\tilde{\gamma}^{\prime }(E_{\zeta |n})\cos ^{2}\zeta ]>0
\end{align*}

Let us study eq. (\ref{Osc1.8.3.3.1}) in more details. The function $\tilde{%
\gamma}(E)$ has the properties: $\tilde{\gamma}(E)=\kappa
_{0}^{-1}|E|^{1/2}+O(|E|^{-1/2})\rightarrow \infty $ as $E\rightarrow
-\infty $; $\tilde{\gamma}(\vartheta _{n}\pm 0)=\pm \infty $, $n\in \mathbb{Z%
}_{+}$; $\tilde{\gamma}(E_{0|n})=0$, $n\in \mathbb{Z}_{+}$; $%
E_{0|n}<\vartheta _{n}<E_{0|n+1}<\vartheta _{n+1}$, $n\in \mathbb{Z}_{+}$.
Then we find: in each energy interval $(\vartheta _{n-1},\vartheta _{n})$, $%
n\in \mathbb{Z}_{+}$, for fixed $\zeta \in (-\pi /2,\pi /2)$, exists one
solution of eq.(\ref{Osc1.8.3.3.1}) $E_{\zeta |n}$ monotonically increasing
from $\vartheta _{n-1}$ through $E_{0|n}$ to $\vartheta _{n}$ when $\zeta $
runs from $-\pi /2+0$ through $0$ to $\pi /2-0$ (we set $\vartheta
_{-1}=-\infty $). Note that the equalities $\lim_{\zeta \rightarrow \pi
/2}E_{\zeta |n}=\lim_{\zeta \rightarrow -\pi /2}E_{\zeta |n+1}=\vartheta
_{n} $ hold which illustrate the equivalence of the extensions with $\zeta
=-\pi /2$ and $\zeta =\pi /2$.

A complete orthonormalized system of (generalized) eigenfunctions of $\hat{h}%
_{\zeta }$ is $\{U_{\zeta |n}(u)=Q_{\zeta |n}U_{\zeta }(u;E_{\zeta
|n}),\;n\in \mathbb{Z}_{+}\}$.

\subsubsection{Spectrum, $\protect\lambda =0$}

Guiding functional, spectral function

\begin{align*}
& U_{\zeta}(u;W)=\frac{\kappa_{0}\sin(W^{1/2}u)}{W^{1/2}}\sin\zeta
+\cos(W^{1/2}u)\cos\zeta, \\
& \sigma_{\zeta}^{\prime}(E)=\frac{1}{\pi}\func{Im}\Omega_{\zeta
}(E+i0),\;\Omega_{\zeta}(W)=\frac{1}{\kappa_{0}}\frac{\kappa_{0}\cos
\zeta+iW^{1/2}\sin\zeta}{\kappa_{0}\sin\zeta-iW^{1/2}\cos\zeta}.
\end{align*}

$\mathbf{a)}\,\zeta=\pm\pi/2$

\begin{align*}
& \Omega_{\pm\pi/2}(W)=i\kappa_{0}^{-2}W^{1/2},\;\Omega_{\pm\pi
/2}(E+i0)=\left\{ 
\begin{array}{c}
i\kappa_{0}^{-2}E^{1/2},\;E\geq0 \\ 
-\kappa_{0}^{-2}|E|^{1/2},\;E<0%
\end{array}
\right. , \\
& \sigma_{\pm\pi/2}^{\prime}(E)=\left\{ 
\begin{array}{c}
\pi^{-1}\kappa_{0}^{-2}E^{1/2},\;E\geq0 \\ 
0,\;E<0%
\end{array}
\right. ,\;\mathrm{spec}\hat{h}_{\pm\pi/2}=[0,\infty)=\mathbb{R}_{+}.
\end{align*}

$\mathbf{b)}\, \zeta=0$

\begin{align*}
& \Omega_{0}(W)=iW^{-1/2},\;\Omega_{0}(E+i0)=\left\{ 
\begin{array}{c}
iE^{-1/2},\;E>0 \\ 
|E|^{-1/2},\;E<0%
\end{array}
\right. , \\
& \sigma_{0}^{\prime}(E)=\left\{ 
\begin{array}{c}
E^{-1/2}/\pi,\;E>0 \\ 
0,\;E<0%
\end{array}
\right. ,\;\mathrm{spec}\hat{h}_{0}=[0,\infty)=\mathbb{R}_{+}.
\end{align*}

{$\mathbf{c)}\,$ General case $|\zeta|<\pi/2$}

In this case, we have%
\begin{equation*}
\sigma_{\zeta}^{\prime}(E)=\frac{1}{\pi\cos^{2}\zeta}\func{Im}\frac {1}{%
\omega_{0}(E+i0)},\;\omega_{0}(W)=-iW^{1/2}+\kappa_{0}\tan\zeta.
\end{equation*}

i) $E\geq0$%
\begin{equation*}
\sigma_{\zeta}^{\prime}(E)=\rho_{\zeta}^{2}(E),\;\rho_{\zeta}(E)=\left( 
\frac{1}{\pi}\frac{E^{1/2}}{\kappa_{0}^{2}\sin^{2}\zeta+E\cos^{2}\zeta }%
\right) ^{1/2}.
\end{equation*}
The spectrum of $\hat{h}_{\zeta}$ is simple and continuous, $\mathrm{spec}%
\hat{h}_{\zeta}=\mathbb{R}_{+}$

ii) $E=-\tau^{2}<0$, $\tau>0$

In this case we have%
\begin{equation*}
\omega _{0}(E)=\tau +\kappa _{0}\tan \zeta .
\end{equation*}

We find:\newline

If $\zeta \in \lbrack 0,\pi /2)$, then $\sigma _{\zeta }^{\prime }(E)=0$ and
there are no spectrum points.

If $\zeta \in (-\pi /2,0)$, then $\func{Im}\Omega _{\zeta }(E+i0)$ can be
different from zero in the point $E_{\zeta |-1}=-\kappa _{0}^{2}\tan
^{2}\zeta $ only and%
\begin{equation*}
\sigma _{\zeta }^{\prime }(E)=Q_{\zeta |-1}^{2}\delta (E-E_{\zeta
|-1}),\;Q_{\zeta |-1}=\sqrt{\frac{2\kappa _{0}\sin |\zeta |}{\cos ^{3}\zeta }%
}.
\end{equation*}

Finally, we obtain%
\begin{equation*}
\mathrm{spec}\hat{h}_{\zeta}=\left\{ 
\begin{array}{l}
\mathbb{R}_{+},\;\zeta\in\lbrack0,\pi/2) \\ 
\mathbb{R}_{+}\cup\{E_{\zeta|-1}\},\;\zeta\in(-\pi/2,0)%
\end{array}
\right. .
\end{equation*}
A complete orthonormalized system of (generalized) eigenfunctions of $\hat{h}%
_{\zeta}$ is

\begin{align*}
& \left\{ 
\begin{array}{c}
\{U_{\zeta|E}(u),\;E\geq0\},,\;\zeta\in\lbrack0,\pi/2) \\ 
\{U_{\zeta|E}(u),\;E\geq0;\;U_{\zeta}(u)\},\;\zeta\in(-\pi/2,0)%
\end{array}
\right. \\
& U_{\zeta|E}(u)=\rho_{\zeta}(E)U_{\zeta}(u;E),\;U_{\zeta}(u)=Q_{\zeta
|-1}U_{\zeta}(u;E_{\zeta|-1}).
\end{align*}

\subsubsection{Spectrum, $\protect\lambda<0$}

First we write down the parameters of the theory in this case:

$\varkappa=e^{-i\pi/4}|\lambda|^{1/4}$, $\varkappa^{-1}=e^{i\pi/4}|%
\lambda|^{-1/4}$, $\varkappa^{2}=-i|\lambda|^{1/2}$,

$\rho=-i|\lambda |^{1/2}u^{2}=e^{-i\pi/2}|\lambda|^{1/2}u^{2}$, $\alpha=1/4-i%
\tilde{w}$, $\tilde{w}\equiv\tilde{w}_{O}=W/4|\lambda|^{1/2}$, ,

$\tilde{\alpha}=1/4+i\tilde{w}=\overline{\alpha_{O}(\overline{\tilde{w}})}%
=1/2-\alpha$, $\tilde{w}(E)=E/4|\lambda|^{1/2}=\overline{\tilde{w}(E)}.$%
\begin{align*}
& \Omega_{\zeta}(W)=-\frac{1}{\kappa_{0}}\frac{\tilde{\gamma}(W)\sin
\zeta-\cos\zeta}{\tilde{\gamma}(W)\cos\zeta+\sin\zeta}, \\
& \tilde{\gamma}(E)=\frac{4\pi|\lambda|^{1/4}}{\kappa_{0}|\Gamma
(\alpha)|^{2}(e^{-\pi\tilde{w}}+ie^{\pi\tilde{w}})}
\end{align*}

$\mathbf{a)}\,\zeta=\pm\pi/2$

\begin{align*}
& \sigma_{O\pm\pi/2}^{\prime}(E)=-\frac{1}{\pi\kappa_{0}}\func{Im}\tilde{%
\gamma}(E+i0)=-\frac{1}{\pi\kappa_{0}}\func{Im}\tilde{\gamma }%
(E)\equiv\rho_{\pm\pi/2}^{2}(E), \\
& \rho_{O\pm\pi/2}(E)=\frac{2|\lambda|^{1/8}e^{\pi\tilde{w}/2}}{\kappa
_{0}|\Gamma(\alpha)|(e^{2\pi\tilde{w}}+e^{-2\pi\tilde{w}})^{1/2}},\;\mathrm{%
spec}\hat{h}_{O\pm\pi/2}=\mathbb{R}.
\end{align*}

$\mathbf{b)}\,\zeta=0$

\begin{align*}
& \sigma_{0}^{\prime}(E)=\frac{1}{\pi\kappa_{0}}\func{Im}\frac {1}{\tilde{%
\gamma}(E+i0)}=\frac{1}{\pi\kappa_{0}}\func{Im}\frac {1}{\tilde{\gamma}(E)}%
\equiv\rho_{0}^{2}(E), \\
& \rho_{0}(E)=\frac{1}{2\pi}e^{\pi\tilde{w}/2}|\Gamma(\alpha)||\lambda
|^{-1/8},\;\mathrm{spec}\hat{h}_{0}=\mathbb{R}.
\end{align*}

$\mathbf{c)}\,$ Arbitrary $\zeta$

\begin{align*}
& \sigma_{\zeta}^{\prime}(E)=\frac{1}{\pi}\func{Im}\Omega_{\zeta }(E+i0)=%
\frac{1}{\pi}\func{Im}\Omega_{\zeta}(E)= \\
& \,=\frac{4|\lambda|^{1/4}}{\kappa_{0}^{2}}\frac{e^{\pi\tilde{w}%
}|\Gamma(\alpha)|^{2}}{e^{2\pi\tilde{w}}|\Gamma(\alpha)|^{4}\sin^{2}\zeta+%
\left( e^{-\pi\tilde{w}}|\Gamma(\alpha)|^{2}\sin\zeta+\frac {%
4\pi|\lambda|^{1/4}}{\kappa_{0}}\cos\zeta\right) ^{2}}\equiv\rho_{%
\zeta}^{2}(E), \\
& \mathrm{spec}\hat{h}_{\zeta}=\mathbb{R}.
\end{align*}

A complete orthonormalized system of (generalized) eigenfunctions of $\hat{h}%
_{\zeta }$ is 
\begin{equation*}
\left\{ U_{\zeta |E}(u)=\rho _{\zeta }(E)U_{\zeta }(u;E),\;E\in \mathbb{R}%
\right\}
\end{equation*}

\subsubsection{S.a. extensions of $\hat{H}_{Os}$}

\begin{equation*}
\mathrm{spec}\hat{H}_{O\zeta _{s}}=\mathrm{spec}\hat{h}_{O\zeta _{s}}.
\end{equation*}%
The eigenfunctions $U_{\hat{H}_{O\zeta _{s}}}(u)$ of the complete set of
eigenfunctions of the operator $\hat{H}_{O\zeta _{s}}$ are equal to 
\begin{equation*}
U_{H_{O\zeta _{s}}}(u)=\frac{1}{\sqrt{2}}U_{h_{O\zeta _{s}}}(|u|).
\end{equation*}

\subsection{S.a. extensions of $\hat{H}_{Oa}$}

Literal consideration gives

\begin{equation*}
\mathrm{spec}\hat{H}_{O\zeta _{a}}=\mathrm{spec}\hat{h}_{O\zeta _{a}},
\end{equation*}%
the eigenfunctions $U_{\hat{H}_{O\zeta _{a}}}(u)$ of the complete set of
eigenfunctions of operator $\hat{H}_{O\zeta _{a}}$ are equal to 
\begin{equation*}
U_{H_{O\zeta _{a}}}(u)=\frac{1}{\sqrt{2}}\varepsilon (u)U_{h_{O\zeta
_{a}}}(|u|).
\end{equation*}

Note that the spectrum of a total s.a. Hamiltonian $\hat{H}_{O\zeta
_{s}\zeta _{a}}=\hat{H}_{O\zeta _{s}}\oplus \hat{H}_{O\zeta _{a}}$ is simple
for $\lambda >0$, $\zeta _{s}\neq \zeta _{a}$, and twofold for $\lambda >0$, 
$\zeta _{s}=\zeta _{a}$, and for $\lambda \leq 0$.

\subsection{Standard extension}

If we consider a differential operation $\check{H}_{O}$ (\ref{Osc1.2.1}) as
acting on complete axis $\mathbb{R}$, a symmetrical operator $\hat{H}_{O(%
\mathbb{R})}$ should be determine as follows:%
\begin{equation*}
\hat{H}_{O(\mathbb{R})}:\left\{ D_{H_{O}(\mathbb{R})}=\mathcal{D}(\mathbb{R}%
),\;\hat{H}_{O(\mathbb{R})}\psi (u)=\check{H}_{O}\psi (u),\;\forall \psi \in
D_{H_{O}(\mathbb{R})}\right\} .
\end{equation*}

Adjoint operator $\hat{H}_{O(\mathbb{R})}^{+}$ is%
\begin{equation*}
\hat{H}_{O(\mathbb{R})}^{+}:\left\{ 
\begin{array}{l}
D_{H_{O(\mathbb{R})}^{+}}=\{\psi _{\ast },\psi _{\ast }^{\prime }\;\mathrm{%
are\;a.c.\;on}\mathcal{\;}\mathbb{R},\;\psi _{\ast },\check{H}_{O}\psi
_{\ast }\in L^{2}(\mathbb{R})\} \\ 
\hat{H}_{O(\mathbb{R})}^{+}\psi _{\ast }(u)=\check{H}_{O}\psi _{\ast
}(u),\;u\in \mathbb{R},\;\forall \psi _{\ast }\in D_{H_{O(\mathbb{R})}^{+}}%
\end{array}%
\right. .
\end{equation*}%
Because $\omega _{H_{O}^{+}}(\psi _{\ast },\chi _{\ast })=\left. \left[ \chi
_{\ast },\psi _{\ast }\right] (u)\right| _{u\rightarrow \infty }-\left. %
\left[ \chi _{\ast },\psi _{\ast }\right] (u)\right| _{u\rightarrow -\infty
}=0$, the operator $\hat{H}_{O(\mathbb{R})}^{+}$ is symmetrical and, as
consequence, s.a.. That means that there is only one s.a. extension of
symmetrical operator $\hat{H}_{O(\mathbb{R})}$, the operator $\hat{H}_{O(%
\mathbb{R})\mathfrak{e}}=\hat{H}_{O(\mathbb{R})}^{+}=\overline{\hat{H}_{O(%
\mathbb{R})}}$, $D_{H_{O(\mathbb{R})\mathfrak{e}}}=D_{H_{O(\mathbb{R})}^{+}}$%
. Because the inclusions%
\begin{equation*}
D_{H_{O(\mathbb{R})}^{+}}\supset \mathcal{D}(\mathbb{R})\supset \mathcal{D}(%
\mathbb{R}\backslash \{0\})
\end{equation*}%
are hold true and $\hat{H}_{O(\mathbb{R})\mathfrak{e}}\mathcal{D}(\mathbb{R}%
\backslash \{0\})=\check{H}_{O}\mathcal{D}(\mathbb{R}\backslash \{0\})=%
\hat{H}_{O}\mathcal{D}(\mathbb{R}\backslash \{0\})$, we obtain that $\hat{H}%
_{O(\mathbb{R})\mathfrak{e}}\supset \hat{H}_{O}$, i. e., $\hat{H}_{O(\mathbb{%
R})\mathfrak{e}}$ is some s.a. extension of symmetrical operator $\hat{H}%
_{O} $. This s.a. extension is specified by the bondary conditions $\psi
_{s}^{\prime }(0)=0$, $\psi _{a}(0)=0$, $\psi _{s,a}\in \left( D_{H_{O(%
\mathbb{R})\mathfrak{e}}}\right) _{s,a}$. thus, we find $\hat{H}_{O(\mathbb{R%
})\mathfrak{e}}=\hat{H}_{O\zeta _{s}\zeta _{a}}$, $\zeta _{s}=0$, $\zeta
_{a}=\pm \pi /2$.

\section{One-dimensional Coulomb-like interaction}

In this section we will consider the equation 
\begin{eqnarray}
&&\,\partial _{x}^{2}\psi (x)+(\frac{3}{16x^{2}}-\frac{g}{|x|}+\mathcal{E}%
)\psi (x)=0,  \label{Coul1.1} \\
&&\mathcal{E}=|\mathcal{E}|e^{i\varphi _{\mathcal{E}}},\;+0\leq \varphi _{%
\mathcal{E}}\leq \pi ,\;\func{Im}\mathcal{E}\geq 0,  \notag
\end{eqnarray}%
where $\hbar ^{2}\mathcal{E}/2m$ is complex energy, $\hbar ^{2}g/2m$ is a
coupling constant. This problem is a particular case of generalized Kratzer
problem, which for has been solved in \cite{BGTV}. Here we will present
those results, which are interesting for investigations of the spectra of
dual theories. Going through the same steps as in Section \ref{Osc}, we find
what follows.

\subsection{Solution on the semiaxis $x>0$}

Introduce a new variable 
\begin{equation*}
z=2Kx,\, K=\sqrt{-\mathcal{E}}=\sqrt{|\mathcal{E} |}e^{i(\varphi_{\mathcal{E}%
}-\pi)/2}=\sqrt{|\mathcal{E}|}\left[ \sin (\varphi_{\mathcal{E}%
}/2)-i\cos(\varphi_{\mathcal{E}}/2)\right] \partial_{x}=2K\partial_{z},\,
\partial_{x}^{2}=4K^{2}\partial_{z}^{2},
\end{equation*}
and new function $\phi(z)=z^{-1/4}e^{z/2}\psi(x).$ Then we obtain

\begin{equation}
z\partial_{z}^{2}\phi(z)+(1/2-z)\partial_{z}\phi(z)-(1/4+g/2K)\phi(z)=0.
\label{Coul1.1.1}
\end{equation}

Eq. (\ref{Coul1.1.1}) is the equation for confluent hypergeometric
functions, in the terms of which we can express solutions of eq. (\ref%
{Coul1.1}). We will use the following solutions:%
\begin{align*}
& C_{+1}(x;\mathcal{E})=\kappa _{0}^{-1/2}x^{3/4}e^{-z/2}\Phi (\alpha
+1/2,3/2;z), \\
& C_{+2}(x;\mathcal{E})=x^{1/4}e^{-z/2}\Phi (\alpha ,1/2;z), \\
& C_{+3}(x;\mathcal{E})=\pi ^{-1/2}\Gamma (\alpha +1/2)x^{1/4}e^{-z/2}\Psi
(\alpha ,1/2;z)= \\
& \,=C_{+2}(x;\mathcal{E})-\frac{2\sqrt{2\kappa _{0}K}\Gamma (\alpha +1/2)}{%
\Gamma (\alpha )}C_{+1}(x;\mathcal{E}),\;\alpha +1/2\neq -n,n\in \mathbb{Z}%
_{+}, \\
& \alpha \equiv \alpha _{C}=1/4+g/2K=1/4-w,\;w\equiv w_{C}=-g/2K.
\end{align*}

\subsubsection{Asymptotics}

Let $x\rightarrow +0$

We have%
\begin{align*}
& C_{+1}(x;\mathcal{E})=\kappa _{0}^{-1/2}x^{3/4}+O(x^{7/4}),\;C_{+2}(x;%
\mathcal{E})=C_{+\mathrm{as}2}(x)+O(x^{9/4}), \\
& C_{+3}(x;\mathcal{E})=C_{+\mathrm{as}2}(x)-\frac{2\sqrt{2K}\Gamma (\alpha
+1/2)}{\Gamma (\alpha )}x^{3/4}+O(x^{7/4}),\;\alpha +1/2\neq -n,\;n\in 
\mathbb{Z}_{+}, \\
& C_{+\mathrm{as}2}(x)=x^{1/4}+2gx^{5/4}.
\end{align*}

Let $x\rightarrow \infty $, $\func{Im}\mathcal{E}>0$ ($\func{Re}K>0$, $-\pi
/2<\arg K\leq 0$, $-\pi /2<\arg z\leq 0$)%
\begin{align*}
& C_{+1}(x;\mathcal{E})=\frac{\sqrt{\pi }(2K)^{\alpha -1}}{2\kappa
_{0}^{1/2}\Gamma (\alpha +1/2)}x^{-w}e^{z/2}(1+O(x^{-1}))=O(x^{-\func{Re}%
w}e^{\func{Re}Kx}), \\
& C_{+2}(x;\mathcal{E})=\frac{\sqrt{\pi }(2K)^{\alpha -1/2}}{\Gamma (\alpha )%
}x^{-w}e^{z/2}(1+O(x^{-1}))=O(x^{-\func{Re}w}e^{\func{Re}Kx}), \\
& C_{+3}(x;\mathcal{E})=\pi ^{-1/2}\Gamma (\alpha +1/2)(2K)^{-\alpha
}x^{w}e^{-z/2}(1+O(x^{-1}))= \\
& =O(x^{\func{Re}w_{C}}e^{-\func{Re}Kx}).
\end{align*}

\subsubsection{The limit $\mathcal{E}=0$}

\begin{align*}
& C_{+1}(x;\mathcal{E})\rightarrow\kappa_{0}^{-1/2}x^{3/4}\Phi
(g/2K,3/2;2Kx)=\frac{x^{1/4}}{2\sqrt{\kappa_{0}g}}\sinh(2\sqrt{gx}), \\
&C_{+1} (x;0)=\frac{x^{1/4}}{2\sqrt{\kappa_{0}g}}\sinh(2\sqrt{gx}); \\
& C_{+2}(x;\mathcal{E})\rightarrow x^{1/4}\Phi(g/2K,1/2;2Kx)=x^{1/4}\cosh(2%
\sqrt {gx}),\;C_{+2}(x;0)=x^{1/4}\cosh(2\sqrt{gx}); \\
& C_{+3}(x;\mathcal{E})\rightarrow x^{1/4}\left[ \cosh(2\sqrt{gx})-\left. 
\frac{\sqrt{2K}\Gamma(\alpha_{C}+1/2)}{\sqrt{g}\Gamma(\alpha_{C})}\right|
_{K\rightarrow0}\sinh(2\sqrt{gx})\right]=x^{1/4}e^{-2\sqrt{gx}}, \\
& C_{+3}(x;0)=x^{1/4}e^{-2\sqrt{gx}}
\end{align*}
(we used a relation 
\begin{equation*}
\left. \sqrt{2K}\Gamma(\alpha_{C}+1/2)/\Gamma(\alpha _{C})\right|
_{K\rightarrow0}=\left. \varkappa\kappa_{0}^{-1/2}\Gamma
(\alpha_{O}+1/2)/\Gamma(\alpha_{O})\right|
_{\varkappa\rightarrow0}=2^{-1}\kappa_{0}^{-1/2}\sqrt{-W}=\sqrt{g}).
\end{equation*}
Thus obtained solutions are in agreement with direct solution of eq. (\ref%
{Coul1.1}) for $\mathcal{E}=0.$

Note, that all solutions of eq.(\ref{Coul1.1}) are square-integrable at the
origin and only the solution $C_{+3}(x;\mathcal{E})$ is square-integrable at
the infinity for $\func{Im}\mathcal{E}>0$, i. e., $C_{+3}(x;\mathcal{E})\in
L^{2}(\mathbb{R}_{+})$ for $\func{Im}\mathcal{E}>0$.

It follows from the relation 9.212.1 of \cite{Grad-Ryzh} that 
\begin{align*}
& e^{-z/2}\Phi (\alpha ,1/2;z)=e^{-Kx}\Phi (1/4+g/2K,1/2;2Kx)= \\
& \,=e^{Kx}\Phi (1/4-g/2K,1/2;-2Kx), \\
& e^{-z/2}\Phi (\alpha +1/2,3/2;z)=e^{-Kx}\Phi (3/4+g/2K,3/2;2Kx)= \\
& \,=e^{Kx}\Phi (3/4-g/2K,3/2;-2Kx),
\end{align*}%
i. e., the functions $C_{+1}$ and $C_{+2}$ are even functions of $K$ ( for
fixed rest parameters and $x$). That means that $C_{+1}$ and $C_{+2}$ are
real-entire functions of $\mathcal{E}$.

The Wronskians of the solutions of eq.(\ref{Coul1.1}) are

\begin{align*}
\mathrm{Wr}(C_{+1},C_{+2}) & =\mathrm{Wr}(C_{+1},C_{+3})=-%
\kappa_{0}^{-1/2}/2, \\
\mathrm{Wr}(C_{+2},C_{+3}) & =-\sqrt{2K}\frac{\Gamma(\alpha+1/2)}{%
\Gamma(\alpha)},\;\left. \mathrm{Wr}(C_{+2},C_{+3})\right| _{\mathcal{E}=0}=-%
\sqrt{g}.
\end{align*}

\subsubsection{Solution on the semiaxis $x<0$}

For $x<0$, we will use the solutions $C_{-k}(x;\mathcal{E})$,%
\begin{equation*}
C_{-k}(x;\mathcal{E})=C_{+k}(|x|;\mathcal{E}),\;k=1,2,3,\;x<0.
\end{equation*}

\subsection{Symmetrical operator $\hat{H}_{C}$}

For given a differential operation $\check{H}_{C}\equiv\check{H}$,%
\begin{equation*}
\check{H}=-\partial_{x}^{2}+\frac{g}{|x|}-\frac{3}{16x^{2}},
\end{equation*}
we determine the following symmetrical operator $\hat{H}_{C}\equiv\hat{H}$,

\begin{equation*}
\hat{H}:\left\{ 
\begin{array}{l}
D_{H}=\mathcal{D}(\mathbb{R}\backslash \{0\}), \\ 
\hat{H}\psi (x)=\check{H}\psi (x),\;\forall \psi \in D_{H},\;%
\end{array}%
\right. .
\end{equation*}

\subsection{Adjoint operator $\hat{H}_{C}^{+}$}

\begin{equation*}
\hat{H}_{C}^{+}\equiv \hat{H}^{+}:\left\{ 
\begin{array}{l}
D_{H^{+}}=\{\psi _{\ast },\psi _{\ast }^{\prime }\;\mathrm{are}\;\mathrm{%
a.c.\;in}\mathcal{\;}\mathbb{R}\backslash \{0\},\;\psi _{\ast },\hat{H}%
_{C}^{+}\psi _{\ast }\in L^{2}(\mathbb{R})\} \\ 
\hat{H}^{+}\psi _{\ast }(x)=\check{H}\psi _{\ast }(x),\;x\in \mathbb{R}%
\backslash \{0\},\;\forall \psi _{\ast }\in D_{H^{+}}%
\end{array}%
\right. .
\end{equation*}

\subsubsection{Asymptotics}

\label{asymp}

I) $|x|\rightarrow\infty$

Because $V(x)=g/|x|-3/16x^{2}\rightarrow 0$ as $|x|\rightarrow \infty $, we
have: $\psi _{\ast },\psi _{\ast }^{\prime }\rightarrow 0$, $\forall \psi
_{\ast }\in D_{H^{+}}$,

$[\psi _{\ast },\chi _{\ast }](x)\rightarrow 0$ $\forall \psi _{\ast },\chi
_{\ast }\in D_{H^{+}}$ as $x\rightarrow \pm \infty $.

II) $x\rightarrow+0$

Because $\check{H}\psi _{\ast }\in L^{2}(\mathbb{R})$, we have%
\begin{equation*}
\check{H}\psi _{\ast }(u)=(-\partial _{x}^{2}+g/|x|-3/16x^{2})\psi _{\ast
}(x)=\eta (x),\;\eta \in L^{2}(\mathbb{R}).
\end{equation*}%
General solution of this equation can be represented in the form%
\begin{align*}
& \psi _{\ast }(x)=a_{+1}C_{+1}(x;0)+a_{+2}C_{+2}(x;0)+I(x), \\
& \psi _{\ast }^{\prime }(x)=a_{+1}C_{+1}^{\prime
}(x;0)+a_{+2}C_{+2}^{\prime }(x;0)+I^{\prime }(x),
\end{align*}%
where%
\begin{align*}
& I(x)=2\sqrt{\kappa _{0}}\left[ C_{+2}(x;0)\int_{0}^{x}C_{+1}(y;0)\eta
(y)dy-C_{+1}(x;0)\int_{0}^{x}C_{+2}(y;0)\eta (y)dy\right] , \\
& I^{\prime }(x)=2\sqrt{\kappa _{0}}\left[ C_{+2}^{\prime
}(x;0)\int_{0}^{x}C_{+1}(y;0)\eta (y)dy-C_{+1}^{\prime
}(x;0)\int_{0}^{x}C_{+2}(y;0)\eta (y)dy\right] .
\end{align*}%
We obtain with the help of the Cauchy-Bunyakovskii inequality:%
\begin{equation*}
I(x)=O(x^{3/2}),\;I^{\prime }(x)=O(x^{1/2}),\;x\rightarrow +0,
\end{equation*}%
so that we have%
\begin{align*}
& \psi _{\ast }(x)=a_{+1}\kappa _{0}^{-1/2}x^{3/4}+a_{+2}C_{+\mathrm{as}%
2}(x)+O(x^{3/2}), \\
& \psi _{\ast }^{\prime }(x)=(3/4)a_{+1}\kappa _{0}^{-1/2}x^{-1/4}+a_{+2}C_{+%
\mathrm{as}2}^{\prime }(x)+O(x^{1/2}).
\end{align*}

III) $x\rightarrow-0$

Analogously, we obtain for $x\rightarrow -0$:%
\begin{align*}
& \psi _{\ast }(x)=a_{-1}\kappa _{0}^{-1/2}|x|^{3/4}+a_{-2}C_{+\mathrm{as}%
2}(|x|)+O(|x|^{3/2}), \\
& \psi _{\ast }^{\prime }(x)=-(3/4)a_{-1}\kappa
_{0}^{-1/2}|x|^{-1/4}-a_{-2}C_{+\mathrm{as}2}^{\prime }(|x|)+O(|x|^{1/2}).
\end{align*}

\subsection{Sesquilinear form $\protect\omega_{H^{+}}(\protect\psi_{\ast},%
\protect\chi_{\ast})$}

\begin{align*}
&
\omega_{H^{+}}(\chi_{\ast},\psi_{\ast})=\omega_{+H^{+}}(\chi_{\ast},\psi_{%
\ast})+\omega_{-H^{+}}(\chi_{\ast},\psi_{\ast}), \\
& \omega_{+H^{+}}(\chi_{\ast},\psi_{\ast})=\int_{0}^{\infty}\left[ \overline{%
\chi_{\ast}(x)}\check{H}\psi_{\ast}(x)-\overline{\check{H}\chi _{\ast}(x)}%
\psi_{\ast}(x)\right] dx= \\
& \,=-\left. \left[ \chi_{\ast},\psi_{\ast}\right] (x)\right|
_{x\rightarrow+0}=\frac{1}{2\kappa_{0}^{1/2}}(\overline{a_{\chi_{\ast}+2}}%
a_{\psi_{\ast}+1}-\overline{a_{\chi_{\ast}+1}}a_{\psi_{\ast}+2}), \\
& \omega_{-H^{+}}(\chi_{\ast},\psi_{\ast})=\int_{-\infty}^{0}\left[ 
\overline{\chi_{\ast}(x)}\check{H}\psi_{\ast}(x)-\overline{\check{H}\chi
_{\ast}(x)}\psi_{\ast}(x)\right] dx= \\
& \,=\left. \left[ \chi_{\ast},\psi_{\ast}\right] (x)\right|
_{x\rightarrow-0}=\frac{1}{2\kappa_{0}^{1/2}}(\overline{a_{\chi_{\ast}-2}}%
a_{\psi_{\ast}-1}-\overline{a_{\chi_{\ast}-1}}a_{\psi_{\ast}-2}),
\end{align*}
such that we have%
\begin{align*}
& \omega_{H^{+}}(\chi_{\ast},\psi_{\ast})=\frac{1}{2\kappa_{0}^{1/2}}(%
\overline{\mathbf{a}_{\chi_{\ast}2}}\mathbf{a}_{\psi_{\ast}1}-\overline {%
\mathbf{a}_{\chi_{\ast}1}}\mathbf{a}_{\psi_{\ast}2})=\frac{i}{4\kappa
_{0}^{3/2}}\left( \overline{\mathbf{b}_{\chi_{\ast}}}\mathbf{b}%
_{\psi_{\ast}}-\overline{\mathbf{d}_{\chi_{\ast}}}\mathbf{d}%
_{\psi_{\ast}}\right) , \\
& \mathbf{a}_{1}=\left( 
\begin{array}{c}
a_{+1} \\ 
a_{-1}%
\end{array}
\right) ,\;\mathbf{a}_{2}=\left( 
\begin{array}{c}
a_{+2} \\ 
a_{-2}%
\end{array}
\right) , \\
& \mathbf{b}=\left( 
\begin{array}{c}
b_{+} \\ 
b_{-}%
\end{array}
\right) =\mathbf{a}_{1}+i\kappa_{0}\mathbf{a}_{2},\;\mathbf{d}=\left( 
\begin{array}{c}
d_{+} \\ 
d_{-}%
\end{array}
\right) =\mathbf{a}_{1}-i\kappa_{0}\mathbf{a}_{2}.
\end{align*}

\subsection{Self-adjoint hamiltonians}

Because all s.a. hamiltonians, $\hat{H}_{C\mathfrak{e}}$, act on its domains
as $\check{H}_{C}$, it should specify definition domains only. The
definition domain $D_{H_{C\mathfrak{e}}}\equiv D_{H_{\mathfrak{e}}}$ of s.a.
operator $\hat{H}_{C\mathfrak{e}}\equiv \hat{H}_{\mathfrak{e}}$ is
determined by condition%
\begin{equation*}
\omega _{H^{+}}(\chi ,\psi )=0,\;\forall \chi ,\psi \in D_{H_{\mathfrak{e}}},
\end{equation*}%
from which it follows%
\begin{equation}
\mathbf{d}_{\psi }=U\mathbf{b}_{\psi },\;\forall \psi \in D_{H_{C\mathfrak{e}%
}},  \label{Coul1.5.1}
\end{equation}%
where $U$ is an arbitrary, but fixed for given extension, unitary $(2\times
2)$-matrix, $U^{+}U=1$. Thus, any s.a. hamiltonian is determined by
assignment of unitary matrix $U$ (we will denote the corresponding s.a.
hamiltonian by $\hat{H}_{CU}$ ($\equiv \hat{H}_{U}$ in this section)),%
\begin{equation*}
\hat{H}_{U}^{+}:\left\{ 
\begin{array}{l}
D_{H_{U}}=\{\psi :\;\psi \in D_{\check{H}}^{\ast },\;\mathbf{d}_{\psi }=U%
\mathbf{b}_{\psi }\} \\ 
\hat{H}_{U}\psi (x)=\check{H}\psi (x),\;x\in \mathbb{R}\backslash
\{0\},\;\forall \psi \in D_{U}%
\end{array}%
\right. .
\end{equation*}%
Thus, there exists a $U(2)$-family of s.a. extensions of the initial
symmetric operator $\hat{H}_{C}$.

\subsection{Parity conserving extensions}

The introduction of the parity opereator is the same as in Section {\ref{Osc}%
}. The $U=U_{P}$ matrix also has the same properties. So we come to defining
the elements of the matrix.

In the terms of the a.b. conditions, the obtained form of the matrix $U_{P}$
means the following:%
\begin{equation}
a_{s,a+1}\cos \zeta _{s,a}=\kappa _{0}a_{s,a+2}\sin \zeta _{s,a},\;|\zeta
_{s,a}|\leq \pi /2,\ \zeta _{s,a}=-\pi /2\sim \zeta _{s,a}=\pi /2,
\label{Coul1.6.4a}
\end{equation}%
or%
\begin{equation}
\psi _{s,a}(x)=\left\{ 
\begin{array}{l}
a[\kappa _{0}^{1/2}x^{3/4}\sin \zeta _{s,a}+C_{+\mathrm{as}2}(x)\cos \zeta
_{s,a}]+O(x^{3/2}),\;x>0 \\ 
\pm a[\kappa _{0}^{1/2}|x|^{3/4}\sin \zeta _{s,a}+C_{+\mathrm{as}2}(|x|)\cos
\zeta _{s,a}]+O(x^{3/2}),\;x<0%
\end{array}%
\right. ,\;x\rightarrow 0,  \label{Coul1.6.4b}
\end{equation}%
where $\zeta _{s,a}=\varphi _{s,a}/2-\pi /2$. The inverse statement is true
as well. Namely, if matrix $U$ gives the boundary condition of the form (\ref%
{Coul1.6.4b}) (or (\ref{Coul1.6.4a})) then that matrix $U$ has the form (\ref%
{Osc1.6.3}) with $\varphi _{s,a}=2\zeta _{s,a}+\pi $. In what follows, we
change the notation of s.a. operator $\hat{H}_{U_{P}}$ for $\hat{H}_{\zeta
_{s,a}}\equiv \hat{H}_{C\zeta _{s,a}}$.

\subsection{Extensions on semiaxis $\mathbb{R}_{+}$}

\subsubsection{Differential operation $\check{h}_{C}$}

\begin{equation*}
\check{h}_{C}\equiv\check{h}=\check{H}=-\partial_{x}^{2}+\frac{g}{|x|}-\frac{%
3}{16x^{2}}.
\end{equation*}

\subsubsection{Symmetrical operator $\hat{h}_{C}$}

\begin{equation*}
\hat{h}_{C}\equiv\hat{h}:\left\{ 
\begin{array}{l}
D_{h_{C}}\equiv D_{h}=\mathcal{D}(\mathbb{R}_{+}) \\ 
\hat{h}\psi(x)=\check{h}\psi(x),\;\forall\psi\in D_{h}%
\end{array}
\right.
\end{equation*}

\subsubsection{Adjoint operator $\hat{h}_{C}^{+}$}

\begin{equation*}
\hat{h}_{C}^{+}\equiv \hat{h}^{+}:\left\{ 
\begin{array}{l}
D_{h^{+}}=\{\psi _{\ast },\psi _{\ast }^{\prime }\;\mathrm{are\;a.c.\;in}%
\mathcal{\;}\mathbb{R}_{+},\;\psi _{\ast },\check{H}\psi _{\ast }\in L^{2}(%
\mathbb{R}_{+})\} \\ 
\hat{h}^{+}\psi _{\ast }(x)=\check{h}\psi _{\ast }(x),\;\forall \psi _{\ast
}\in D_{h^{+}}%
\end{array}%
\right. .
\end{equation*}

\subsubsection{Asymptotics of $\protect\psi _{\ast }\in D_{h^{+}}$}

Literally repeating the considerations of \ref{asymp} we obtain:

I) $x\rightarrow\infty$

$[\psi_{\ast},\chi_{\ast}](x)\rightarrow0$ as $x\rightarrow\infty$, $%
\forall\psi_{\ast},\chi_{\ast}\in D_{h^{+}}$.

II) $x\rightarrow 0$%
\begin{align*}
& \psi _{\ast }(x)=a_{+1}\kappa _{0}^{-1/2}x^{3/4}+a_{+2}C_{+\mathrm{as}%
2}(x)+O(x^{3/2}), \\
& \psi _{\ast }^{\prime }(x)=(3/4)a_{+1}\kappa _{0}^{-1/2}x^{-1/4}+a_{+2}C_{+%
\mathrm{as}2}^{\prime }(x)+O(x^{1/2}).
\end{align*}

\subsection{Sesquilinear form $\protect\omega_{h^{+}}(\protect\psi_{\ast},%
\protect\chi_{\ast})$}

\begin{align*}
& \omega_{h^{+}}(\chi_{\ast},\psi_{\ast})=\int_{0}^{\infty}\left[ \overline{%
\chi_{\ast}(x)}\check{h}\psi_{\ast}(x)-\overline{\check{h}\chi _{\ast}(x)}%
\psi_{\ast}(x)\right] dx= \\
& \,=-\left. \left[ \chi_{\ast},\psi_{\ast}\right] (x)\right|
_{x\rightarrow0}=\frac{1}{2\kappa_{0}^{1/2}}(\overline{a_{\chi_{\ast}+2}}%
a_{\psi_{\ast}+1}-\overline{a_{\chi_{\ast}+1}}a_{\psi_{\ast}+2})= \\
& =\frac{i}{4\kappa_{0}^{3/2}}\left( \overline{b_{\chi_{\ast}}}b_{\psi
_{\ast}}-\overline{d_{\chi_{\ast}}}d_{\psi_{\ast}}\right)
,\;b=a_{1}+i\kappa_{0}a_{2},\;d=a_{1}-i\kappa_{0}a_{2}.
\end{align*}

\subsection{Self-adjoint hamiltonians}

Because all s.a. hamiltonians, $\hat{h}_{C\mathfrak{e}}\equiv\hat{h}_{%
\mathfrak{e}}$, act on their domains as $\check{h}$, we should specify only
definition domains. The definition domain $D_{h_{\mathfrak{e}}}$ of s.a.
operator $\hat{h}_{\mathfrak{e}}$ is determined by condition%
\begin{equation*}
\omega_{h^{+}}(\chi,\psi)=0,\;\forall\chi,\psi\in D_{h_{\mathfrak{e}}},
\end{equation*}
from which it follows%
\begin{equation*}
d_{\psi}=e^{i\varphi}b_{\psi},\;\forall\psi\in D_{h_{C\mathfrak{e}%
}},\;0\leq\varphi\leq2\pi,\;0\sim2\pi,
\end{equation*}
or, equivalently%
\begin{equation*}
a_{1}\cos\zeta=\kappa_{0}a_{2}\sin\zeta,\;\zeta=\varphi/2-\pi/2,\;-\pi
/2\sim\pi/2.
\end{equation*}
Thus, any s.a. hamiltonian is determined by assignment of unitary matrix $%
U(1)=e^{i\varphi}$ (we will denote the corresponding s.a. hamiltonian by $%
\hat{h}_{C\zeta}\equiv\hat{h}_{\zeta}$),%
\begin{equation*}
\hat{h}_{\zeta}:\left\{ 
\begin{array}{l}
D_{h_{\zeta}}\equiv\{\psi:\;\psi\in D_{\check{h}}^{\ast},\;a_{1}\cos
\zeta=\kappa_{0}a_{2}\sin\zeta\} \\ 
\hat{h}_{\zeta}\psi(x)=\check{h}\psi(x),\;\forall\psi\in D_{h_{\zeta}}%
\end{array}
\right. .
\end{equation*}
Equivalently, the boundary condition for $\psi\in D_{h_{\zeta}}$ can be
represented in the form%
\begin{equation}
\psi(x)=a[\kappa_{0}^{1/2}x^{3/4}\sin\zeta+C_{+\mathrm{as}2}(x)\cos
\zeta]+O(x^{3/2})\rightarrow0.  \label{Coul1.7.5.1}
\end{equation}
Thus, there exists a $U(1)$-family of s.a. extensions $\hat{h}_{\zeta}$ of
the initial symmetric operator $\hat{h}$.

\subsection{Self-adjoint extensions of $\hat{H}_{s}$}

The Hilbert space $L_{s}^{2}(\mathbb{R})$ is the space of all symmetric
functions that are square integrable on $\mathbb{R}$. For these functions,
the relations 
\begin{equation}
(\chi ,\psi )=2(\chi ,\psi )_{+}\ ,\ \omega _{H^{+}}(\chi ,\psi )=2\omega
_{H^{+}}(\chi ,\psi )_{+}=2\omega _{h^{+}}(\chi ,\psi ),  \label{Coul1.8.1}
\end{equation}%
hold true, where%
\begin{equation}
(\chi ,\psi )_{+}=\int_{0}^{\infty }\overline{\chi (x)}\psi \left( x\right)
dx,  \label{Coul1.8.2}
\end{equation}%
and $\omega _{H^{+}}(\chi ,\psi )_{+}=\omega _{h^{+}}(\chi ,\psi )$ is the
sesquilinear form with respect to the scalar product (\ref{Coul1.8.2}).

Let us consider the isometry $T$: $\psi\in\mathbb{R}\overset{T}{%
\longrightarrow}\sqrt{2}\psi$, $\psi\in\mathbb{R}_{+}$. Then%
\begin{equation}
D_{H_{s}}\overset{T}{\longrightarrow}D_{h}=\mathcal{D}(\mathbb{R}_{+}),\ \
D_{H_{s}^{+}}=D_{\check{H}}^{\ast}\left( \mathbb{R}\right) \overset{T}{%
\longrightarrow}D_{h^{+}}=D_{\check{h}}^{\ast}\left( \mathbb{R}_{+}\right) .
\label{Coul1.8.3}
\end{equation}

It follows from eqs. (\ref{Coul1.8.1}) and (\ref{Coul1.8.3}) that there is
one-to-one correspondence (the isometry $T$) between s.a. extensions $\hat{H}%
_{\zeta_{s}}$ of the symmetric operator $\hat{H}_{s}$ in $L_{s}^{2}(\mathbb{R%
})$ and s.a. extensions $\hat{h}_{\zeta}$ of the symmetric operator $\hat{h}$
in $L^{2}(\mathbb{R}_{+})$: $\hat{H}_{\zeta_{s}}\overset{T}{%
\Longleftrightarrow}\hat{h}_{\zeta}$, $\zeta_{s}=\zeta$. Thus, the spectral
analysis of s.a. operator $\hat{H}_{\zeta_{s}}$ in $L_{s}^{2}(\mathbb{R})$
is reduced to the spectral analysis of s.a. operator $\hat{h}_{\zeta}$, $%
\zeta_{s}=\zeta$, in $L^{2}(\mathbb{R}_{+})$. Below, we represent this
analysis.

\subsubsection{Green function $G_{C\protect\zeta }(x,y;\mathcal{E})$,
spectral function $\protect\sigma _{C\protect\zeta }(E)$}

We find the Green function $G_{C\zeta }(x,y;\mathcal{E})\equiv G_{\zeta
}(x,y;\mathcal{E})$ as the kernel of the integral representation 
\begin{equation*}
\psi (x)=\int_{0}^{\infty }G_{\zeta }(x,y;\mathcal{E})\eta (y)dy,\;\eta \in
L^{2}(\mathbb{R}_{+}),
\end{equation*}%
of unique solution of an equation%
\begin{equation}
(\hat{h}_{\zeta }^{+}-\mathcal{E})\psi (x)=\eta (x),\;\func{Im}\mathcal{E}>0,
\label{Coul1.8.2.1}
\end{equation}%
for $\psi \in \in D_{h_{\zeta }}$, that is, $\psi \in L^{2}(\mathbb{R}_{+})$
and $\psi (x)$ satisfies the boundary conditions (\ref{Coul1.7.5.1}). We
find 
\begin{align}
& G_{\zeta }(x,y;\mathcal{E})=-\Omega _{C\zeta }(\mathcal{E})U_{\zeta }(x;%
\mathcal{E})U_{\zeta }(y;\mathcal{E})-2\kappa _{0}^{-1/2}\left\{ 
\begin{array}{c}
\tilde{U}_{\zeta }(x;\mathcal{E})U_{\zeta }(y;\mathcal{E}),\;x>v \\ 
U_{\zeta }(x;\mathcal{E})\tilde{U}_{\zeta }(y;\mathcal{E}),\;x<y%
\end{array}%
\right. ,  \label{Coul1.8.2.2} \\
& \Omega _{C\zeta }(\mathcal{E})\equiv \Omega _{\zeta }(\mathcal{E})=\frac{2%
\tilde{\omega}_{\zeta }(\mathcal{E})}{\kappa _{0}^{1/2}\omega _{\zeta }(%
\mathcal{E})},\;\omega _{C\zeta }(\mathcal{E})\equiv \omega _{\zeta }(%
\mathcal{E})=\tilde{\gamma}(\mathcal{E})\cos \zeta +\sin \zeta ,  \notag \\
& \tilde{\omega}_{C\zeta }(\mathcal{E})\equiv \tilde{\omega}_{\zeta }(%
\mathcal{E})=\tilde{\gamma}(\mathcal{E})\sin \zeta -\cos \zeta ,\;\tilde{%
\gamma}(\mathcal{E})=2\sqrt{\frac{2K}{\kappa _{0}}}\gamma ({\alpha }%
),\,\gamma ({\alpha })=\frac{\Gamma ({\alpha +1/2})}{\Gamma ({\alpha })} , 
\notag
\end{align}%
where we used relations%
\begin{align*}
& U_{C\zeta }(x;\mathcal{E})\equiv U_{\zeta }(x;\mathcal{E})=\kappa
_{0}C_{+1}(x;\mathcal{E})\sin \zeta +C_{+2}(x;\mathcal{E})\cos \zeta , \\
& \tilde{U}_{C\zeta }(x;\mathcal{E})\equiv \tilde{U}_{\zeta }(x;\mathcal{E}%
)=\kappa _{0}C_{+1}(x;\mathcal{E})\cos \zeta -C_{+2}(x;\mathcal{E})\sin
\zeta , \\
& C_{+3}(x;W)=U_{\zeta }(x;W)\tilde{\omega}_{\zeta }(W)-\tilde{U}_{\zeta
}(x;W)\omega _{\zeta }(W).
\end{align*}%
Note that $U_{\zeta }(x;W)$ and $\tilde{U}_{\zeta }(x;\mathcal{E})$ are
real-entire solutions of eq. (\ref{Coul1.1}), $U_{\zeta }(x;W)$ satisfies
the boundary conditions (\ref{Coul1.7.5.1}), and the last term in the r.h.s.
of eq. (\ref{Coul1.8.2.2}) is real for $\mathcal{E}=E$ ($\func{Im}\mathcal{E}%
=0$).

\subsubsection{Guiding functional}

The guiding functional $\Phi_{C\zeta}(\xi;W)\equiv\Phi_{\zeta}(\xi;W)$ is%
\begin{align*}
& \Phi_{\zeta}(\xi;\mathcal{E})=\int_{0}^{\infty}U_{C\zeta}(x;\mathcal{E}%
)\xi(x)dx,\;\xi\in\mathbb{D}_{\zeta}=D_{h_{C\zeta}}\cap D_{r}(\mathbb{R}%
_{+}), \\
& U_{C\zeta}(x;\mathcal{E})\equiv U_{\zeta}(x;\mathcal{E})=%
\kappa_{0}C_{+1}(x;\mathcal{E})\sin\zeta+C_{+2}(x;\mathcal{E})\cos\zeta.
\end{align*}

The guiding functional $\Phi _{\zeta }(\xi ;\mathcal{E})$ is simple and the
spectrum of $\hat{h}_{C\zeta }$ is simple. From the relation%
\begin{equation*}
U_{\zeta }^{2}(x_{0};E)\sigma _{\zeta }^{\prime }(E)=\frac{1}{\pi }\func{Im}%
G_{\zeta }(x_{0}-0,x_{0}+0;E+i0),
\end{equation*}%
we find%
\begin{equation*}
\sigma _{\zeta }^{\prime }(E)=-\frac{1}{\pi }\func{Im}\Omega _{\zeta }(E+i0).
\end{equation*}

\subsubsection{Spectrum, $E\geq0$}

\label{Eless}

We have $\varphi_{\mathcal{E}}=+0$, $K=-i\sqrt{E}$, $\alpha=1/4-i\tilde{w}$, 
$\tilde{w}=-g/2\sqrt{E}$,%
\begin{align*}
& \tilde{\gamma}(E)=2\sqrt{\frac{2}{\kappa_{0}}}e^{-i\pi/4}E^{1/4}\frac{%
\Gamma(1-(1/4+i\tilde{w}))\Gamma(1/4+i\tilde{w})}{|\Gamma(\alpha)|^{2}}= \\
& =\frac{4\sqrt{2}\pi E^{1/4}(e^{-\pi\tilde{w}_{C}}-ie^{\pi\tilde{w}_{C}})}{%
\kappa_{0}^{1/2}|\Gamma(\alpha_{C})|^{2}(e^{2\pi\tilde{w}_{C}}+e^{-2\pi%
\tilde{w}_{C}})}.
\end{align*}

First we will study a question whether there exists the eigenvalue $E=0$.
Linearly independent solutions of eq. (\ref{Coul1.1}) for $\mathcal{E}=0$ are%
\begin{equation*}
C_{+1}(x;0)=\frac{x^{1/4}}{2\sqrt{\kappa_{0}g}}\sinh(2\sqrt{gx}%
),\;C_{+3}(x;0)=x^{1/4}e^{-2\sqrt{gx}}.
\end{equation*}
For $g\leq0$, the square-integrable solutions are absent. For $g>0$, there
is one square-integrable solution $C_{+3}(x;0)$. Representing the asymptotic
of the function $C_{+3}(x;0)$ in the form%
\begin{equation*}
C_{+3}(x;0)=(-2\kappa_{0}^{-1/2}g^{1/2})\kappa_{0}^{1/2}x^{3/4}+C_{2\mathrm{%
as}}(x)+O(x^{7/4}),\;x\rightarrow0,
\end{equation*}
we find that this function is eigenfunction of s.a. hamiltonian $\hat{h}%
_{\zeta_{g}}$, $\zeta_{g}=\arctan(-2\kappa_{0}^{-1/2}g^{1/2})$.

{$\mathbf{a)}$\thinspace $\zeta ={\pm }\pi /2$}

\begin{align*}
& -\frac{1}{\pi }\func{Im}\Omega _{\pm \pi /2}(E)=-\frac{1}{\pi }\func{Im}%
\frac{2}{\kappa _{0}^{1/2}}\tilde{\gamma}(E)=\frac{8\sqrt{2}E^{1/4}e^{\pi 
\tilde{w}}}{\kappa _{0}|\Gamma (\alpha )|^{2}(e^{2\pi \tilde{w}}+e^{-2\pi 
\tilde{w}})}\equiv \\
& \,\equiv \rho _{C\pm \pi /2}^{2}(E)\equiv \rho _{\pm \pi
/2}^{2}(E),\;\lim_{E\rightarrow +0}\rho _{\pm \pi /2}^{2}(E)=\left\{ 
\begin{array}{c}
0,\;g\geq 0 \\ 
4|g|^{1/2}/\pi \kappa _{0},\;g<0%
\end{array}%
\right. , \\
& \sigma _{\pm \pi /2}^{\prime }(E)=\rho _{\pm \pi /2}^{2}(E),\;\mathrm{spec}%
\hat{h}_{\pm \pi /2}=\mathbb{R}_{+}.
\end{align*}

{$\mathbf{b)}$\thinspace\ $\zeta =0$}%
\begin{align*}
& -\frac{1}{\pi }\func{Im}\Omega _{0}(E)=\frac{1}{\pi }\func{Im}\frac{2}{%
\kappa _{0}^{1/2}\tilde{\gamma}(E)}=\frac{e^{\pi \tilde{w}_{C}}|\Gamma
(\alpha )|^{2}}{2\sqrt{2}\pi ^{2}E^{1/4}}\equiv \rho _{C0}^{2}(E)\equiv \rho
_{0}^{2}(E), \\
& \lim_{E\rightarrow +0}\rho _{0}^{2}(E)=\left\{ 
\begin{array}{l}
0,\;g>0 \\ 
1/\pi |g|^{1/2},\;g<0%
\end{array}%
\right. , \\
& \sigma _{0}^{\prime }(E)=\rho _{0}^{2}(E),\;\mathrm{spec}\hat{h}_{0}=%
\mathbb{R}_{+}.
\end{align*}

{$\mathbf{c)}$\thinspace\ $\zeta \neq 0,\pm \pi /2$, $E>0$}%
\begin{align*}
& -\frac{1}{\pi }\func{Im}\Omega _{\zeta }(E)=\frac{2}{\pi \kappa _{0}^{1/2}}%
\func{Im}\frac{e^{-\pi \tilde{w}}\cos \zeta -a\sin \zeta +ie^{\pi \tilde{w}%
}\cos \zeta }{e^{-\pi \tilde{w}}\sin \zeta +a\cos \zeta +ie^{\pi \tilde{w}%
}\sin \zeta }= \\
& \,=\frac{8\sqrt{2}E^{1/4}e^{\pi \tilde{w}}|\Gamma (\alpha )|^{2}}{(\kappa
_{0}^{1/2}e^{-\pi \tilde{w}}|\Gamma (\alpha )|^{2}\sin \zeta +4\sqrt{2}\pi
E^{1/4}\cos \zeta )^{2}+\kappa _{0}e^{2\pi \tilde{w}}|\Gamma (\alpha
)|^{4}\sin ^{2}\zeta }\equiv \\
& \equiv \rho _{C\zeta }^{2}(E)\equiv \rho _{\zeta }^{2}(E),\;a=\frac{4\sqrt{%
2}\pi E^{1/4}}{\kappa _{0}^{1/2}|\Gamma (\alpha )|^{2}}.
\end{align*}

Now we can calculate $\lim_{E\rightarrow+0}\rho_{\zeta}^{2}(E)$.

i) $g\leq0$%
\begin{equation*}
\lim_{E\rightarrow+0}\rho_{\zeta}^{2}(E)=4\pi^{-1}|g|^{1/2}(4|g|\cos^{2}%
\zeta+\kappa_{0}\sin^{2}\zeta)^{-1}.
\end{equation*}

ii) $g>0$, $\zeta \neq \zeta _{g}$%
\begin{equation*}
\lim_{E\rightarrow +0}\rho _{\zeta }^{2}(E)=0.
\end{equation*}

iii) $g>0$, $\zeta=\zeta_{g}$%
\begin{align*}
& \sigma_{\zeta_{g}}^{\prime}(E)=16g^{3/2}(1+4g/\kappa_{0})\delta
(E)+\rho_{\zeta_{g}}^{2}(E),\;\rho_{\zeta_{g}}^{2}(0)=\lim_{E\rightarrow
+0}\rho_{\zeta_{g}}^{2}(E)=0, \\
& \mathrm{spec}\hat{h}_{\zeta}=\mathbb{R}_{+}.
\end{align*}

Finally, we obtain for $E\geq 0$:%
\begin{equation*}
\sigma _{\zeta }^{\prime }(E)=\left\{ 
\begin{array}{l}
\rho _{\zeta }^{2}(E),\;\zeta \neq \zeta _{g} \\ 
16g^{3/2}(1+4g/\kappa _{0})\delta (E)+\rho _{\zeta _{g}}^{2}(E),\;\zeta
=\zeta _{g}%
\end{array}%
\right. .
\end{equation*}

Of course, the expressions of subsubsecs \ref{Eless}.a and \ref{Eless}.b are
the limiting cases of the expressions of subsubsec \ref{Eless}.c.

\subsubsection{Spectrum, $E<0$,}

In this case we have

$\varphi_{\mathcal{E}}=\pi-0$, $K(E)=K|_{W=E}=\sqrt{|E|}$, $\alpha=1/4-w$, $%
w(E)=w|_{\mathcal{E}=E}=-g/2\sqrt{|E|}$, $\tilde{\gamma}(E)=2\sqrt{2}%
\kappa_{0}^{-1/2}|E|^{1/4}\Gamma(\alpha+1/2)/\Gamma(\alpha)$

{$\mathbf{a)}$\thinspace\ $\zeta =\pi /2$}

We find%
\begin{equation*}
-\Omega_{\pm\pi/2}(\mathcal{E}) =-\frac{2\tilde{\gamma}(\mathcal{E})}{%
\kappa_{0}^{1/2}}=-\frac{4\sqrt{2}|E|^{1/4}\Gamma(\alpha+1/2)}{%
\kappa_{0}\Gamma(\alpha)}.
\end{equation*}

i) $g\geq0$

In this case, $\Omega _{\pi /2}(E)$ is real and finite, such that $\func{Im}%
\Omega _{\pi /2}(E)=\sigma _{\pm \pi /2}^{\prime }(E)=0$ and $\mathrm{spec}%
\hat{h}_{\pi /2}=\varnothing $.

ii) $g<0$, $w(E)=|g|/2\sqrt{|E|}$.

In this case, the function $\Omega _{\pi /2}(E)$ is real for $E$ when $%
|\Omega _{\pi /2}(E)|<\infty $. Therefore, $\func{Im}\Omega _{\pi /2}(E+i0)$
can be not equal to zero only in the point $\Omega _{\pi /2}(E)=\pm \infty $%
, i. e., in the points $\alpha =\alpha _{\pm \pi /2|n}=-1/2-n$, $w=w_{\pm
\pi /2|n}=n+3/4$, $E=\vartheta _{Cn}\equiv \vartheta _{n}$, $n\in \mathbb{Z}%
_{+}$, 
\begin{equation*}
\vartheta _{n}=-g^{2}[(2n+1)+1/2]^{-2}.
\end{equation*}%
In the neighborhood of the points $\vartheta _{n}$ we have ($\mathcal{E}%
=\vartheta _{n}+\Delta $, $\Delta =E-\vartheta _{n}+i\varepsilon $, $\alpha (%
\mathcal{E})=-1/2-n-b\Delta $, $b=|g||\vartheta _{n}|^{-3/2}/4$)%
\begin{align*}
& -\func{Im}\Omega _{\pi /2}(E+i0)=-\frac{4\sqrt{2}|\vartheta _{n}|^{1/4}}{%
\kappa _{0}\Gamma (-1/2-n)}\func{Im}\Gamma (-n-b\Delta )|_{\varepsilon
\rightarrow +0}= \\
& \,=\pi Q_{\pi /2|n}^{2}\delta (E-\vartheta _{n}),\;Q_{\pi /2|n}=\left[ 
\frac{8\sqrt{2}|\vartheta _{n}|^{7/4}(2n+1)!!}{\sqrt{\pi }\kappa _{0}|g|2n!!}%
\right] ^{1/2}.
\end{align*}%
Finally, we find%
\begin{equation*}
\sigma _{\pi /2}^{\prime }(E)=\sum_{n=0}^{\infty }Q_{\pi /2|n}^{2}\delta
(E-\vartheta _{n}),\;\mathrm{spec}\hat{h}_{\pi /2}=\{\vartheta _{n}\;n\in 
\mathbb{Z}_{+}\}.
\end{equation*}

We obtain the same results for the case $\zeta =-\pi /2$.

{$\mathbf{b)}$\,$\zeta=0$}

In this case, we have%
\begin{equation*}
-\Omega_{0}(\mathcal{E})=\frac{2}{\kappa_{0}^{1/2}\tilde{\gamma}(\mathcal{E})%
}=\frac{\Gamma(\alpha)}{\sqrt{2}K^{1/2}\Gamma(\alpha+1/2)},\;-\Omega _{0}(E)=%
\frac{\Gamma(\alpha)}{\sqrt{2}|E|^{1/4}\Gamma(\alpha+1/2)}.
\end{equation*}

i) $g\geq0$

In this case, $\Omega_{0}(E)$ is real and finite, such that $\func{Im}%
\Omega_{0}(E)=\sigma_{0}^{\prime}(E)=0$ and $\mathrm{spec}\hat{h}%
_{0}=\varnothing$.

ii) $g<0$, $w(E)=|g|/2\sqrt{|E|}$.

In this case, the function $\Omega _{0}(E)$ is real for $E$ when $|\Omega
_{0}(E)|<\infty $. Therefore, $\func{Im}\Omega _{0}(E+i0)$ can be not equal
to zero only in the point $\Omega _{0}(E)=\pm \infty $, i.e., in the points $%
\alpha =\alpha _{0|n}=-n$, $w=w_{0|n}=n+1/4$, $E=E_{0|n}$, $n=0,1,2,...$,%
\begin{equation*}
|E_{0|n}|^{1/2}=|g|(2n+1/2)^{-1},\;E_{0|n}=-g^{2}(2n+1/2)^{-2}.
\end{equation*}%
In the neighborhood of the points $E_{0|n}$ we have ($\mathcal{E}%
=E_{0|n}+\Delta $, $\Delta =E-E_{0|n}+i\varepsilon $, $\alpha (\mathcal{E}%
)=-n-b\Delta $, $b=|g||E_{0|n}|^{-3/2}/4$)%
\begin{align*}
& -\func{Im}\Omega _{0}(E+i0)=\frac{1}{\sqrt{2}|E_{0|n}|^{1/4}\Gamma (1/2-n)}%
\func{Im}\Gamma (-n-b\Delta )|_{\varepsilon \rightarrow +0}= \\
& \,=\pi Q_{0|n}^{2}\delta (E-E_{0|n}),\;Q_{0|n}=\left[ \frac{2\sqrt{2}%
|E_{0|n}|^{5/4}(2n-1)!!}{\sqrt{\pi }|g|(2n)!!}\right] ^{1/2}.
\end{align*}%
Finally, we find%
\begin{equation*}
\sigma _{0}^{\prime }(E)=\sum_{n=0}^{\infty }Q_{0|n}^{2}\delta (E-E_{0|n}),\;%
\mathrm{spec}\hat{h}_{0}=\{E_{0|n,}\;n\in \mathbb{Z}_{+}\}.
\end{equation*}

{$\mathbf{c)}$\, General case $|\zeta|<\pi/2$}

In this case, we have%
\begin{equation*}
\sigma _{\zeta }^{\prime }(E)=\frac{2}{\pi \kappa _{0}^{1/2}\cos ^{2}\zeta }%
\func{Im}\frac{1}{\tilde{\gamma}(E+i0)+\tan \zeta }
\end{equation*}

The function $\tilde{\gamma}(E)$ is real. Therefore, $\sigma_{\zeta}^{\prime
}(E)$ can be not equal to zero only in the points, where 
\begin{equation}
\tilde{\gamma}(E_{\zeta|n})=-\tan\zeta,  \label{Coul1.8.4.3.1}
\end{equation}
such that we have%
\begin{align*}
\sigma_{\zeta}^{\prime}(E) & =\sum_{n}Q_{\zeta|n}^{2}\delta(E-E_{\zeta
|n}),\;Q_{\zeta|n}=\left( -\frac{2}{\kappa_{0}^{1/2}\cos^{2}\zeta}\frac {1}{%
\tilde{\gamma}^{\prime}(E_{\zeta|n})}\right) ^{1/2}, \\
\tilde{\gamma}^{\prime}(E_{\zeta|n}) & <0,\;\partial_{\zeta}E_{\zeta
|n}=-1/[\cos^{2}\zeta\tilde{\gamma}^{\prime}(E_{\zeta|n})]>0.
\end{align*}

Let us study eq. (\ref{Coul1.8.4.3.1}) in more details.

i) $g\geq0$,

In this case, we have $w(E)\leq0$; $\tilde{\gamma}(E)>0$; $\tilde{\gamma }%
(E)=2^{3/2}\kappa_{0}^{-1/2}\Gamma^{-1}(1/4)%
\Gamma(3/4)|E|^{1/4}+O(|E|^{-1/4})\rightarrow\infty$ as $E\rightarrow-\infty$%
. Eq. (\ref{Coul1.8.4.3.1}) has no solutions.for $\zeta\in(\zeta_{0},\pi/2)$
and for any fixed $\zeta\in(-\pi/2,\zeta_{0})$ has one solution $%
E_{\zeta}^{(-)}\in(-\infty,0)$ monotonically increasing from $-\infty$ to $%
-0 $ as $\zeta$ run from $-\pi/2+0$ to $-0$ (let us remind that for $g>0$
and $\zeta=\zeta_{0}$, there exists the level $E_{\zeta_{0}}^{(-)}=0$).

ii) $g<0$, $w(E)=|g|/2\sqrt{|E|}$

In this case, we have: $\tilde{\gamma}(E)=2^{3/2}\kappa _{0}^{-1/2}\Gamma
^{-1}(1/4)\Gamma (3/4)|E|^{1/4}+O(|E|^{-1/4})$ as $E\rightarrow -\infty $; $%
\tilde{\gamma}(E_{0|n})=0$; $\tilde{\gamma}(\vartheta _{n}\pm 0)=\pm \infty $%
; $E_{0|n}<\vartheta _{n}<E_{0|n+1}<\vartheta _{n+1}$. Then, in any domain $%
(\vartheta _{n-1},\vartheta _{n})$, $n\in \mathbb{Z}_{+}$, for fixed $\zeta
\in (-\pi /2,\pi /2)$, eq. (\ref{Coul1.8.4.3.1}) has one solution $E_{\zeta
|n}$ monotonically increasing from $\vartheta _{n-1}+0$ through $E_{0|n}$ to 
$\vartheta _{n}-0$ as $\zeta $ run from $-\pi /2+0$ through $0$ to $\pi /2-0$
(we set $\vartheta _{-1}=-\infty $)

\section{Comparison of the spectra of the theories of oscillator and
Coulomb- like potential (1D Anyon)}

Making the identifications%
\begin{align}
& u=\sqrt{x/\kappa _{0}},\;W=-4\kappa _{0}g,\;\lambda =-4\kappa _{0}^{2}%
\mathcal{E},  \notag \\
& x=\kappa _{0}u^{2},\;\mathcal{E}=-\lambda /4\kappa _{0}^{2},\;g=-W/4\kappa
_{0},\;\Rightarrow  \label{Cor}
\end{align}%
we will get the following correspondence between oscillator and coulomb
parameters and functions 
\begin{align*}
& K=\sqrt{\lambda }/2\kappa _{0}=\varkappa ^{2}/2\kappa _{0},\;\sqrt{K}%
=\varkappa /\sqrt{2\kappa _{0}},\;z=\rho ,\;\sqrt{g}=\sqrt{-W}/2\sqrt{\kappa
_{0}}, \\
& \alpha _{C}=\alpha _{O},\;w_{C}=w_{O},\;\tilde{w}_{C}=\tilde{w}%
_{O}\;(E_{C}>0,\;\lambda <0;\;see\;8.3), \\
& \gamma _{C}(\alpha _{C})=\gamma _{O}(\alpha _{O}),\;\omega _{C\zeta }(%
\mathcal{E})=\omega _{O\zeta }(W),\;\tilde{\omega}_{C\zeta }(\mathcal{E})=%
\tilde{\omega}_{O\zeta }(W), \\
& \Omega _{C\zeta }(\mathcal{E})=2\kappa _{0}^{1/2}\Omega _{O\zeta }(W).
\end{align*}

Then we'll obtain%
\begin{align*}
C_{+k}(x;\mathcal{E})& =x^{1/4}O_{+k}(u;W),\;k=1,2,3, \\
C_{+\mathrm{as}1}(x;\mathcal{E})& =x^{1/4}O_{+\mathrm{as}}(u;W),
\end{align*}%
in agreement with eq.\ref{eq5}.

It's easy to see, that for any fixed $\zeta$, to each point of continuous
spectrum in the plane $E_{C}, g$ corresponds a point of continuous spectrum
in the plane $E_{O},\lambda$, and to each point of discrete spectrum in the
plane $E_{C}, g$ corresponds a point of discrete spectrum in plane $%
E_{O},\lambda$, while the image of the point, which is not a spectrum point
in the plane $E_{C}, g$, is not point a spectrum point in the plane $%
E_{O},\lambda$, and visa versa. Note, that a complete correspondence between
the points of the spectra exists only if one takes into accounta
``nonphysical'' $\lambda<0$ in the case of oscillator.

The general statement on correspondence of the spectra of two problems is
easily checked in the cases of $\zeta = \pm \pi/2$ and $\zeta=0$.

It is worth mentioning, that as was stated in previous sections, the
complete orthonormalized system of (generalized) eigenfunctions of theories
are $U_{\zeta |E}(u)=\rho _{\zeta }(E)U_{\zeta }(u;E)$ for continuous
spectrum and $U_{\zeta |n}(u)=Q_{\zeta |n}U_{\zeta }(u;E)$ for discrete
spectra. The connections between the normalized functions in two cases are%
\begin{equation*}
U{_{C\zeta |E_{C}}}\left( x{;}g\right) =\frac{\rho {{_{C\zeta }}\left( {E}%
_{C}{,g}\right) }}{\rho {{_{O\zeta }}\left( {E}_{O},\lambda \right) }}{%
x^{1/4}}U_{O\zeta |E_{O}}(u;\lambda )=\sqrt{\frac{|u|}{2}}U_{O\zeta
|E_{O}}(u;\lambda ),
\end{equation*}%
\begin{equation*}
U{_{C\zeta |n}}\left( x{;}g\right) =\frac{{{Q_{C\zeta |n}}\left( g\right) }}{%
{{Q_{O\zeta |n}}\left( \lambda \right) }}{x^{1/4}}U_{O\zeta |n}(u;\lambda ).
\end{equation*}%
Note that the construction of the theory in the way described in this
article automatically produces normalized wave functions. For the descrete
spectrum for (in our terminology for standart extension $\zeta _{s}=0$, $%
\zeta _{a}=\pm \pi /2$ ) we obtain the connection between the
oscillator-anyon wave functions derived in \TEXTsymbol{\backslash}%
cite\{Ter-Ant\} . \cite{Ter-Ant}.

\begin{acknowledgement}
I.T. thanks RFBR Grand 08--01-00737 for partial support.
\end{acknowledgement}


\begin{thebibliography}{9}
\bibitem{Ter-Ant} V. Ter-Antonyan, \emph{\ Dyon-Oscillator Duality}
arXiv:quant-ph/0003106

\bibitem{Ners-Ter-Ant} A.Nersessian, V.M.Ter-Antonyan, \emph{Anyons,
Monopole and Coulomb Problem}, Phys.Atom.Nucl. 61 (1998) 1756-1761

\bibitem{Hak-Ter-Ant} Ye. Hakobyan, V. Ter-Antonyan, \emph{Quantum
oscillator as 1D anyon}, arXiv:quant-ph/0002069

\bibitem{Grad-Ryzh} I.S. Gradshteyn, I.M. Ryzhik. \emph{Tables of integrals,
series, and products} (5ed., Academic Press, 1996)(1762s)

\bibitem{Naima} M.A. Naimark,, \emph{Part II: Linear differential operators
in Hilbert space}, Frederick Ungar Publishing Co., New York, 1968

\bibitem{AkhGlaz} N.I. Akhiezer and LM. Glazman, \emph{Theory of Linear
Operators in Hilbert Space} (Pitman, Boston 1981) N.I. Akhiezer and LM.

Glazman, \emph{Theory of Linear Operators in Hilbert Space}(Nauka,Moscow
1966) (in Russian)

\bibitem{BGTV} M.C. Baldiotti, D.M. Gitman, I.V. Tyutin, B.L. Voronov, \emph{%
Self-adjoint extensions and spectral analysis in the generalized Kratzer
problem}, arXiv:1009.4903
\end{thebibliography}
\end{document}